\documentclass[iop, twocolappendix, numberedappendix, tighten, appendixfloats]{emulateapj}
\usepackage[stable]{footmisc}
\usepackage{graphicx}
\usepackage{txfonts}
\usepackage{natbib}
\bibpunct[, ]{(}{)}{;}{a}{}{,}

\newcommand{\farc}{\hbox{$.\!\!^{\prime\prime}$}} 
 
\newcommand{\erg}{$\rm{erg\,cm^{-2}\,s^{-1}}$} 
\newcommand{\kms}{$\rm{km\,s^{-1}}\,$}
\newcommand{\hb}{H$\beta$} 
\newcommand{\ha}{H$\alpha$}
 
\newcommand{\hi}{\mbox{H\,{\sc i}}}

\newcommand{\oii}{[\ion{O}{2}]} 
\newcommand{\oiii}{[\ion{O}{3}]}
\newcommand{\nii}{[\ion{N}{2}]} 

\newcommand{\griz}{$g' r' i' z'$}
\newcommand{\JHK}{$JHK_{\rm{s}}$}

\newcommand{\Msun}{$M_\odot$}

\begin{document}

\title{\vspace{-0.5cm}Molecular Hydrogen in the Damped Lyman$\alpha$ System towards GRB~120815A at $\lowercase{z}=2.36^{\dag}$}
\altaffiltext{$^{\dag}$}{Based on observations collected at the European Southern Observatory, Paranal, Chile, Program ID: 089.A-0067(B).}

\author{T.~Kr\"{u}hler\altaffilmark{1}, 
C.~Ledoux\altaffilmark{2}, 
J.~P.~U. Fynbo\altaffilmark{1}, 
P.~M.~Vreeswijk\altaffilmark{3}, 
S.~Schmidl\altaffilmark{4},
D.~Malesani\altaffilmark{1}, 
L.~Christensen\altaffilmark{1},
A.~De~Cia\altaffilmark{3}, 
J.~Hjorth\altaffilmark{1}, 
P.~Jakobsson\altaffilmark{5}, 
D.~A.~Kann\altaffilmark{4}, 
L.~Kaper\altaffilmark{6}, 
S.~D.~Vergani\altaffilmark{7, 8},
P.~M.~J.~Afonso\altaffilmark{9},
S.~Covino\altaffilmark{10}, 
A.~de~Ugarte~Postigo\altaffilmark{11}, 
V.~D'Elia\altaffilmark{7, 12}, 
R.~Filgas\altaffilmark{13}, 
P.~Goldoni\altaffilmark{14}, 
J.~Greiner\altaffilmark{15}, 
O.~E.~Hartoog\altaffilmark{6},
B.~Milvang-Jensen\altaffilmark{1}, 
M.~Nardini\altaffilmark{16}, 
S.~Piranomonte \altaffilmark{7},
A.~Rossi\altaffilmark{4}, 
R.~S\'anchez-Ram\'irez\altaffilmark{11},
P.~Schady\altaffilmark{15}, 
S.~Schulze\altaffilmark{17, 18}, 
V.~Sudilovsky\altaffilmark{15}, 
N.~R.~Tanvir\altaffilmark{19},
G.~Tagliaferri\altaffilmark{7},
D.~J.~Watson\altaffilmark{1},
K.~Wiersema\altaffilmark{19},
R.~A.~M.~J.~Wijers\altaffilmark{6}}
\author{D.~Xu\altaffilmark{1}}

\altaffiltext{1}{Dark Cosmology Centre, Niels Bohr Institute, University of Copenhagen, Juliane Maries Vej 30, 2100 K\o benhavn \O, Denmark}\altaffiltext{2}{European Southern Observatory, Alonso de C\'{o}rdova 3107, Vitacura, Casilla 19001, Santiago 19, Chile}\altaffiltext{3}{Department of Particle Physics and Astrophysics, Faculty of Physics, Weizmann Institute of Science, Rehovot 76100, Israel}\altaffiltext{4}{Th\"uringer Landessternwarte Tautenburg, Sternwarte 5, 07778 Tautenburg, Germany}\altaffiltext{5}{Centre for Astrophysics and Cosmology, Science Institute, University of Iceland, Dunhagi 5, IS-107 Reykjavik, Iceland}\altaffiltext{6}{Astronomical Institute Anton Pannekoek, University of Amsterdam, Science Park 904, NL-1098 XH Amsterdam, the Netherlands}\altaffiltext{7}{INAF-Osservatorio Astronomico di Roma, Via Frascati 33, I-00040 Monteporzio Catone, Italy}\altaffiltext{8}{GEPI-Observatoire de Paris, CNRS UMR 8111, Univ. Paris-Diderot, 5 Place Jules Janssen - 92190 Meudon, France}\altaffiltext{9}{American River College, Physics and Astronomy Dpt., 4700 College Oak Drive, Sacramento, CA 95841, USA}\altaffiltext{10}{INAF, Osservatorio Astronomico di Brera, Via E. Bianchi 46, I-23807 Merate, Italy}\altaffiltext{11}{Instituto de Astrof\'{\i}sica de Andaluc\'{\i}a (IAA-CSIC), Glorieta de la Astronom\'{\i}a s/n, 18008, Granada, Spain}\altaffiltext{12}{ASI-Science Data Centre, Via Galileo Galilei, I-00044 Frascati, Italy}\altaffiltext{13}{Institute of Experimental and Applied Physics, Czech Technical University in Prague, Horska 3a/22, 128 00 Prague 2, Czech Republic}
\altaffiltext{14}{APC, Astroparticules et Cosmologie, Universite Paris Diderot, CNRS/
IN2P3, CEA/Irfu, Observatoire de Paris, Sorbonne Paris Cite, 10, Rue Alice Domon et L
eonie Duquet, 75205 Paris Cedex 13, France}\altaffiltext{15}{Max-Planck-Institut f\"{u}r extraterrestrische Physik, Giessenbachs
tra\ss e, 85748 Garching, Germany}
\altaffiltext{16}{Universit\`a degli studi di Milano-Bicocca, Piazza della Scienza 3,
 20126, Milano, Italy}
\altaffiltext{17}{Pontificia Universidad Cat\'{o}lica de Chile, Departamento de Astro
nom\'{\i}a y Astrof\'{\i}sica, Casilla 306, Santiago 22, Chile}\altaffiltext{18}{Millennium Center for Supernova Science}\altaffiltext{19}{Department of Physics and Astronomy, University ofLeicester, University Road, Leicester, LE1 7RH, UK}\

\begin{abstract}
{We present the discovery of molecular hydrogen (H$_2$), including the presence of vibrationally-excited H$_2^{*}$ in the optical spectrum of the afterglow of GRB~120815A at $z=2.36$ obtained with X-shooter at the VLT. Simultaneous photometric broad-band data from GROND and X-ray observations by \textit{Swift}/XRT place further constraints on the amount and nature of dust along the sightline. The galactic environment of GRB~120815A is characterized by a strong DLA with $\log(N\rm{(\hi)/cm^{-2}}) = 21.95\pm0.10$, prominent H$_2$ absorption in the Lyman-Werner bands ($\log(N\rm{(H_2)/cm^{-2}}) = 20.54\pm0.13$) and thus a molecular gas fraction $\log f(\rm{H}_2)=-1.14\pm0.15$. The distance $d$ between the absorbing neutral gas and GRB~120815A is constrained via photo-excitation modeling of fine-structure and meta-stable transitions of \ion{Fe}{2} and \ion{Ni}{2} to $d=0.5\pm0.1$~kpc.  The DLA metallicity ($\rm{[Zn/H]} = -1.15 \pm 0.12$), visual extinction ($A_V\lesssim0.15$~mag) and dust depletion ($\rm{[Zn/Fe]} = 1.01 \pm 0.10$) are intermediate between the values of well-studied, H$_2$-deficient GRB-DLAs observed at high spectral resolution, and the approximately solar metallicity, highly-obscured and H$_2$-rich GRB~080607 sightline. With respect to $N(\hi)$, metallicity, as well as dust-extinction and depletion, GRB~120815A is fairly representative of the average properties of GRB-DLAs. This demonstrates that molecular hydrogen is present in at least a fraction of the more typical GRB-DLAs, and H$_2$ and H$_2^*$ are probably more wide-spread among GRB-selected systems than the few examples of previous detections would suggest. Because H$_2^*$ transitions are located redwards of the Lyman$\alpha$ absorption, H$_2^*$ opens a second route for positive searches for molecular absorption also in GRB afterglows at lower redshifts and observed at lower spectral resolution. Further detections of molecular gas in GRB-DLAs would allow statistical studies, and, coupled with host follow-up and sub-mm spectroscopy, provide unprecedented insights into the process and conditions of star-formation at high redshift.}
\end{abstract}


\section{Introduction}

The formation of stars is tightly correlated with the presence of molecular hydrogen H$_2$ \citep[e.g.,][]{1987ARA&A..25...23S, 1993prpl.conf..125B, 2007ARA&A..45..565M, 2008AJ....136.2846B}. In the local Universe, for example, star-formation proceeds quite exclusively in cold molecular clouds, and molecular gas is ubiquitous in the interior of the Galaxy \citep[e.g.,][]{1977ApJ...216..291S, 2001ApJ...547..792D} or the Magellanic Clouds \citep[e.g.,][]{2002ApJ...566..857T}. Similarly, at higher redshift, the increased star-formation rates of non-merging, $z \sim 1-2$ galaxies are likely related to an elevated gas-mass fraction \citep[e.g.,][]{2010Natur.463..781T, 2010ApJ...713..686D}. Direct detections of H$_2$ beyond $z\sim0$ are, however, challenging from the ground, and its presence and state at high redshift are most commonly inferred indirectly via tracers such as CO or HCN. The fact that these molecules have dipole moments makes them much more amenable to detection via mm/sub-mm spectroscopy, but inferences about H$_2$ then rely on poorly-known conversion factors. 

Because the absorption signatures of H$_2$, the Lyman-Werner bands, are located bluewards of the Lyman$\alpha$ transition, H$_2$ is most directly probed at $z\gtrsim 2$ through high-resolution spectroscopy\footnote{High spectral resolution is required to disentangle H$_2$ features from the Ly$\alpha$ forest.} of damped Lyman$\alpha$ (DLA) systems infront of quasi-stellar objects (QSOs) or $\gamma$-ray bursts (GRBs) \citep[e.g.,][]{2003MNRAS.346..209L}. But even when probed in very high detail with QSO-DLAs, detections of molecular gas are scarce: H$_2$, for example, was detected only along 13 out of 77 QSO sightlines in the sample of \citet{2008A&A...481..327N}, while observations of other molecules like HD \citep[e.g.,][]{2001AstL...27..683V, 2010MNRAS.403.1541M, 2011A&A...526L...7N} or CO \citep[e.g.,][]{2008A&A...482L..39S, 2009A&A...503..765N, 2011A&A...526L...7N} remain limited to very few systems at high redshift.

{A noteworthy difference between the DLAs of long\footnote{GRBs are typically divided into long and short events with different progenitor channels. In this work, we focus exclusively on long GRBs, typically defined as having durations $T_{90}  > 2\,\rm{s}$ \citep[see, e.g.,][for details]{1993ApJ...413L.101K}. $T_{90}$ is the time during which 90\% of the GRB's $\gamma$-ray photons are emitted.} GRBs and QSOs lies within their different physical nature and selection methods. QSO-DLAs are intervening systems discovered in optical spectra of e.g., color-selected QSOs because of their large \hi~cross section \citep[e.g.,][]{2005ApJ...635..123P, 2012A&A...547L...1N}. In contrast, GRB-DLAs are located inside the star-forming host galaxy and are initially localized via the GRB's $\gamma$-ray and/or X-ray emission \citep[e.g.,][and references therein]{2008ApJ...683..321F}.  The GRB's high energy emission is virtually unaffected by dust obscuration. Afterglows can further be, albeit for a short time, much more luminous in the optical wavelength range \citep[e.g.,][]{2009ApJ...691..723B} than QSOs before they fade rapidly on timescales of hours or days. A proxy for the environments which GRB-DLAs probe is the distribution of visual extinctions derived from afterglow data for an unbiased sample of \textit{Swift} GRBs: $1/2$ of all GRBs are located behind a dust column with $A_V \lesssim 0.2\,\rm{mag}$, $\approx$ 30\% show $0.2\,\rm{mag} < A_V \lesssim 1\,\rm{mag}$, and 20\% of all \textit{Swift} GRB have an $A_V > 1$~mag \citep{2011A&A...526A..30G, 2013MNRAS.tmp.1196C}. GRBs offer thus the potential to probe environments from diffuse to translucent gas clouds.}

Molecular gas remained elusive for a long time in GRB-DLAs \citep[e.g.,][]{2007ApJ...668..667T, 2009A&A...506..661L}, and has now been unequivocally detected only in the low-resolution spectra of the afterglows of one \citep[GRB~080607,][]{2009ApJ...691L..27P} and possibly a second \citep[GRB~060206,][]{2006A&A...451L..47F, 2008A&A...489...37T} GRB-DLA. Given the very high column densities of gas regularly detected in GRB afterglow spectra \citep[e.g.,][]{2001A&A...370..909J, 2003ApJ...597..699H, 2004A&A...419..927V, 2006A&A...460L..13J} and their direct link to star-formation \citep[e.g.,][]{1998Natur.395..670G, 2003Natur.423..847H}, this low detection rate of H$_2$ was initially quite surprising. Because the material probed by GRB-DLAs is in many cases located at distances of several hundred parsecs or more from the $\gamma$-ray burst \citep[e.g.,][]{2007A&A...468...83V, 2011A&A...532C...3V, 2009ApJ...694..332D, 2009ApJ...701L..63S}, the afterglow's UV flux itself will not photo-dissociate the H$_2$ of the GRB-DLA \citep[e.g.,][]{2007ApJ...668..667T, 2009A&A...506..661L}. A genuine deficit of molecular hydrogen in GRB-DLAs as compared to QSO-DLAs for example, would instead indicate that H$_2$ is absent already prior to the burst. This could be caused by substantial background UV-radiation fields in their host galaxies, which are capable of dissociating hydrogen molecules efficiently \citep[e.g.,][]{2008ApJ...682.1114W}.

The incidence of H$_2$ along QSO sightlines \citep[e.g.,][]{1985MNRAS.212..517L, 1998A&A...335...33S, 2000A&A...364L..26P} is positively correlated with metallicity and dust depletion \citep{2003MNRAS.346..209L, 2006A&A...456L...9P, 2008A&A...481..327N}. Dust absorbs photons in the rest-frame ultra-violet (UV) very efficiently. {In UV-steep reddening laws similar to the one observed towards the Small Magellanic Cloud (SMC) and often along GRB sightlines, for example, the ratio between the extinction in the Lyman-Werner bands and visual extinction $A_{1000\,\rm{\AA}}/A_V$ exceeds a factor of 6. Even small values of $A_V$ thus provide considerable suppression of the observed flux in the rest-frame UV.} The small sample ($\sim$10--15 events) of GRB afterglows observed with echelle spectrographs with a very high resolving power ($R\gtrsim 40\,000$) is thus restricted to the most luminous and least obscured end of the afterglow brightness distribution. Consequently, the best studied GRB-selected absorbers are typically those with comparatively low dust and metal content \citep{2007ApJS..168..231P, 2009ApJ...694..332D, 2010A&A...523A..36D, 2012A&A...545A..64D}. {Given their low metal abundance and dust depletion, the previous non-detections of H$_2$ in GRB-DLAs observed at high spectral resolution are thus consistent with QSO-DLA number statistics \citep{2009A&A...506..661L}.}

The full sample of GRB afterglows, including those observed with lower spectral resolution \citep[e.g.,][]{2009ApJS..185..526F}, however, is diverse and includes direct evidence for metal-rich gas \citep[e.g.,][]{2006ApJ...652.1011W, 2007ApJ...666..267P, 2009ApJ...697.1725E, 2011arXiv1110.4642S}. In addition significant dust-depletion factors \citep[e.g.,][]{2004ApJ...614..293S, 2013MNRAS.428.3590T, 2013arXiv1301.3912H}, large amounts of reddening \citep[e.g.,][]{2011A&A...526A..30G, 2012ApJ...753...82Z} and strong metal lines \citep[e.g.,][]{2003ApJ...585..638S, 2011ApJ...727...73C, 2012A&A...548A..11D} are regularly detected towards GRBs. 

{Most GRB-DLAs indeed show $\log(N\rm{(Fe)^{dust}/cm^{-2}}) >14.7$ \citep{2013arXiv1305.1153D}, which is the critical value above which molecular hydrogen has been observed towards QSO \citep{2008A&A...481..327N}. GRB-DLAs are thus promising sight-lines to detect and analyze molecules and their associated gas at high redshift.} The detection of H$_2$, however, has proven to be difficult for GRB-DLAs: {observationally, it is much easier to disentangle H$_2$ transitions from the Ly$\alpha$ forest with high spectral resolution and signal-to-noise ratio (S/N) in the rest-frame UV, favoring bright targets. Physically, however, a significant column density of H$_2$ is present in environments that are likely metal-rich and dust-depleted (and are thus probed by more obscured, UV-faint afterglows).}

In fact, it required the exceptional event of GRB~080607 to clearly detect both H$_2$ and CO along a GRB sightline. GRB~080607 had one of the intrinsically most luminous afterglows ever discovered \citep{2010arXiv1009.0004P}, and was observed spectroscopically only minutes after the initial trigger when the transient was still bright \citep{2008GCN..7849....1P}. The GRB~080607 afterglow spectrum is characterized by large amounts of neutral gas ($\log(N\rm{(\hi)/cm^{-2}}) \sim 22.7$) around solar metallicity, a very high dust content of $A_V \sim 3.3$~mag and evidence for a 2175~\AA~feature \citep{2009ApJ...691L..27P, 2010arXiv1009.0004P}. It further exhibits molecular absorption of H$_2$ and CO \citep{2009ApJ...691L..27P} including vibrationally-excited H$_2^*$ \citep{2009ApJ...701L..63S} at $z=3.04${, the first report of H$_2^*$ absorption at high redshift. Excited H$_2$ absorption lines are rarely detected towards bright stars in the Galaxy \citep{1995ApJ...445..325F, 2001ApJ...553L..59M}, but more commonly in emission in IR spectra of the central galaxies of galaxy (proto)-clusters \citep[e.g.,][]{2000ApJ...545..670D, 2007MNRAS.382.1246J, 2012ApJ...751...13O} or active galactic nuclei \citep[e.g.,][]{1978ApJ...222L..49T, 1997ApJ...477..631V, 2012ApJ...747...95G}.}

In this work, we present the detection of molecular hydrogen in the DLA of GRB~120815A at $z = 2.36$. Its afterglow was observed with X-shooter in the spectral range from the UV-cutoff to $2.5\,\mu\rm{m}$ at medium resolving power ($R\sim 5000-10000$), providing both the S/N and spectral sampling necessary to investigate the physical conditions in the GRB-DLA. Our spectroscopic data are complemented by simultaneous optical/near-infrared (NIR) photometry and X-ray observations, providing strong constraints on the reddening along the sightline. 

Throughout the paper we quote errors at the 1$\sigma$ confidence level, report magnitudes in the AB system, and provide wavelengths and redshifts in a vacuum, heliocentric reference frame. The atomic data used in this work are those of \citet{2003ApJS..149..205M} with few exceptions\footnote{Atomic data for \ion{Ni}{2}($\lambda\,1370$) are from \citet{2006ApJ...637..548J} and \ion{Ni}{2}$\,^4\rm{F}_{9/2}$ as well as \ion{Fe}{2}$\,^4\rm{F}_{9/2}$ from \texttt{http://www.cfa.harvard.edu/amp/ampdata/kurucz23/sekur.html}. We note that we apply a similar modification to \ion{Ni}{2}$\,^4\rm{F}_{9/2}$ as in \citet{2007A&A...468...83V}. In the photo-excitation modelling, we also use atomic data provided by \citet{1999ApJS..120..101V} and references therein.}. Solar reference abundances were taken from Table 1 of \citet{2009ARA&A..47..481A} and follow the suggestions of \citet{2009LanB...4B...44L} whether abundances from the solar photosphere, meteorites or the average between the two values are used.

\section{Observations and Data Reduction}
\subsection{\textit{Swift} Observations}
\label{Swift}

The Burst Alert Telescope (BAT; \citealp{2005SSRv..120..143B}) onboard the \textit{Swift} satellite \citep{2004ApJ...611.1005G} triggered on the long and soft GRB~120815A on 2012-08-15 at $T_0 =$~02:13:58~UTC \citep{2012GCN..13645...1P, 2012GCN..13652...1M}. \textit{Swift}'s two narrow-field X-ray and UV/optical instruments, the X-ray Telescope (XRT; \citealp{2005SSRv..120..165B}) and Ultra-Violet Optical Telescope (UVOT; \citealp{2005SSRv..120...95R}) started observing the GRB field approximately 2.7~ks after the trigger, and detected the fading afterglow associated with GRB~120815A \citep{2012GCN..13647...1K, 2012GCN..13666...1H}. X-ray data were retrieved from the XRT online repository \citep{2007A&A...469..379E, 2009MNRAS.397.1177E}, and analyzed in \texttt{XSPEC v12.7.0} \citep{1996ASPC..101...17A}. Due to the faintness of the afterglow at the time of the observation, imaging data from UVOT have limited S/N, are not constraining and are thus not used in the further analysis.

\subsection{GROND Optical/NIR Photometry}
\label{grond}

The Gamma-Ray burst Optical/NIR Detector (GROND; \citealp{2007Msngr.130...12G, 2008PASP..120..405G}) initiated automatic follow-up observations of GRB~120815A on 2012-08-15 02:14:24 UTC, with the first data taken $\sim$150~s after $T_0$ \citep{2012GCN..13648...1S}. Simultaneous imaging in four optical (\griz) and three NIR ($JHK_{\rm{s}}$) filters was performed continuously until 05:47 UTC. {Our analysis of the GROND data focuses on the broad-band spectral energy distribution (SED) and absolute flux scale contemporaneously with the X-ray (Section~\ref{Swift}) and optical/NIR spectroscopy (Section \ref{xs}), as well as the afterglow's UV-photon yield for the modeling of the excited absorption lines (Section~\ref{sec:distance}). A detailed discussion and theoretical interpretation of the afterglow multi-color light curve is beyond the scope of this work.}

GROND data were reduced and analysed within pyraf/IRAF \citep{1993ASPC...52..173T} in a standard manner \citep{2008ApJ...685..376K}. The transient is detected with high S/N in each filter. Using USNO field stars as astrometric reference, we derive a position for GRB~120815A of R.A.~(J2000)~=~18:15:49.83, Dec.~(J2000)~=~$-$52:07:52.5 with an absolute accuracy of 0\farc{3} in each coordinate. The photometric solution was tied to the magnitudes of stars from the SDSS catalog \citep{2011ApJS..193...29A} in \griz~ observed directly before and after the GRB field, and 2MASS field stars \citep{2006AJ....131.1163S} in \JHK. Based on the scatter of individual calibration stars we estimate our absolute photometric accuracy to be 4\% in \griz, 5\% in $J$ and $H$ and 8\% in $K_{\rm{s}}$, dominating the total photometric error. The GROND data are reported in Tables~\ref{tab:grizphot} and \ref{tab:JHKphot}, and the multi-color light curves that were used as input to the photo-excitation modeling are shown in Figure~\ref{fig:lc}.

\subsection{X-shooter Optical/NIR Spectroscopy}
\label{xs}

Spectroscopic observations of the GRB~120815A afterglow in the wavelength range between $3000$~and $24\,800\,$\AA~commenced on 2012-08-15 at 03:55 UT (6.06~ks after the BAT trigger) with the cross-dispersed echelle spectrograph X-shooter \citep{2011arXiv1110.1944V} mounted at ESO's Very Large Telescope (VLT) UT2. They consisted of four nodded exposures in the sequence ABBA with exposure times of $600\,$s each, taken simultaneously in X-shooter's ultraviolet/blue (UVB), visible (VIS) and near-infrared (NIR) arms. The average airmass was $1.37$ and the mid-time of the stacked exposure is 04:17:50 UTC (7.43~ks after the BAT trigger or 37 minutes in the rest-frame at $z=2.36$). Sky conditions were clear and dark with a median seeing of 0\farc{6} during the observations. X-shooter spectroscopy was performed with slit-widths of 1\farc{0}, 0\farc{9} and 0\farc{9} in the UVB, VIS and NIR arm, respectively. The resolving power $R = \lambda/\Delta\lambda$ was derived from unsaturated, single telluric lines in the red arms, and scaled to the UVB arm using a standard dependence of seeing with wavelength \citep[see][for details]{2011MNRAS.413.2481F} and tabulated values\footnote{\texttt{http://www.eso.org/sci/facilities/paranal/\\instruments/xshooter/}} of $R$ for different X-shooter slit widths. This results in values of $R \sim 6000$, 10400, 6200 for the UVB/VIS and NIR arm, respectively. Because of the good seeing, this is somewhat better than the nominal values for the given slit widths derived from arclamp frames ($R = 5380$, 8800, 5300).

X-shooter data were reduced with the ESO/X-shooter pipeline \texttt{v1.5.0} \citep{2006SPIE.6269E..80G}, rectifying the data on an output grid with a dispersion of $0.15$~\AA~per pixel in the UVB, $0.13~$\AA/pixel in the VIS and $0.5~$\AA/pixel in the NIR arm, respectively. The wavelength-to-pixel conversion was chosen as a compromise between slightly oversampling X-shooter's spectral resolution, while at the same time limiting the correlation of adjacent pixels in the rectification step. The wavelength solution was obtained against arc-lamp frames in each arm, leaving residuals with a root mean square of 0.03/0.04/0.3~\AA, respectively, corresponding to 3/2/5~km~s$^{-1}$ at 4000/7500/$16\,500\,$\AA. Given the projected proximity of the GRB to the USNO star 0378-0977505 and to maximize S/N, each of the four frames in the individual arms was reduced separately, with the sky estimated in regions free of source counts. The sky region is limited because of the short slit length of X-shooter ($11{\arcsec}$) and artifacts at the order edges, affecting the accuracy and quality of the sky-subtraction in the NIR arm.

The spectra were extracted from the combined frame in a variance-weighted method, taking into account the wavelength-dependence of the centroid and width of the trace. To estimate the error in the combined spectrum, we used the noise model of the X-shooter pipeline for each of the four individual frames propagated through the stacking and extraction process. The median S/N per resolution element $\Delta\lambda$ is $\sim15$ ($4200 - 5700$~\AA), $\sim16$ ($5700-10\,000$~\AA),  and $\sim$9/14/16/15  in $Y$/$J$/$H$/$K$.

Flux-calibration was performed against the spectro-photometric standard LTT7987 observed starting on 2012-08-15 at 05:00 UTC. The atmospheric dispersion correctors of X-shooter in the UVB and VIS arm were switched off after a technical problem. Because of the substantial airmass during our observations we expect some chromatic slit-loss in the data that we aim to correct via our afterglow model (see Section~\ref{agmodel}). {The flux-calibrated X-shooter spectrum was fitted with a low-order polynomial (iteratively excluding the absorption lines), which yields the continuum level of the data. The ratio between continuum fit and afterglow model then provides wavelength-dependent matching factors over the whole spectral range of X-shooter, and thus an accurately flux-calibrated spectrum.}

\section{Results}

\subsection{Broad-band Spectral Energy Distribution}
\label{agmodel}

\begin{figure}
\includegraphics[angle=0, width=0.96\columnwidth]{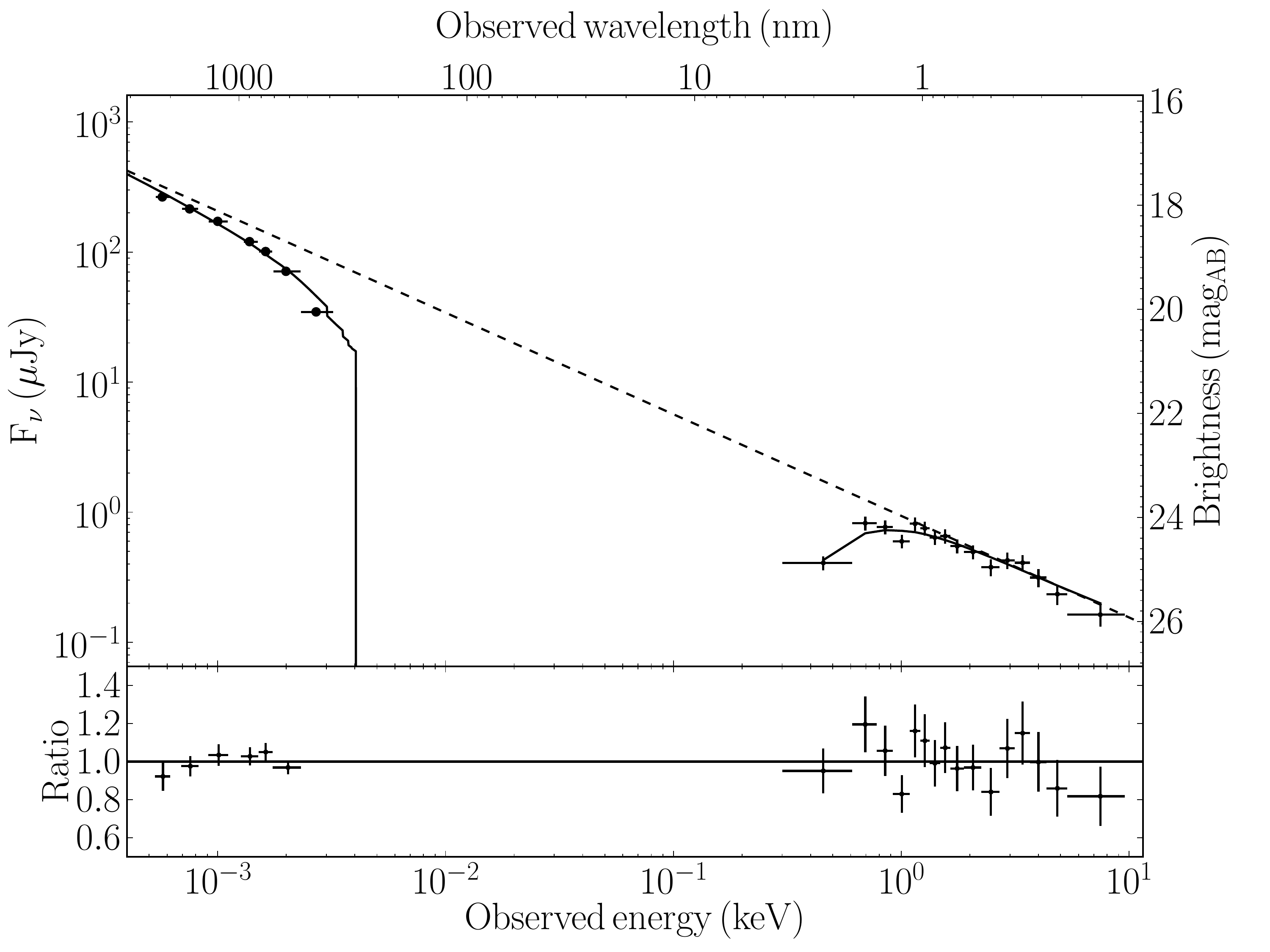}
\caption{NIR-to-X-ray spectral energy distribution and model for the afterglow of GRB~120815A at 7.3~ks after the trigger. Solid lines show the model including gas and dust absorption, while the dashed line illustrates the underlying synchrotron continuum emission. X-ray data have been binned to yield a S/N of at least 8 to enhance clarity. The $g'$-band photometry is not fitted, because this filter extends bluewards of the Ly$\alpha$ transition.}
\label{fig:bbsed}
\end{figure}

We fitted the broad-band afterglow data from XRT and GROND under the assumption that the underlying afterglow flux is well represented by synchrotron emission and the afterglow is in the slow-cooling regime \citep[see, e.g.,][]{2002ApJ...568..820G} with the optical/X-ray energy range being above the afterglow's characteristic synchrotron frequency. Single and broken power-law models were used, where the soft X-ray column density is attributed to two absorbers with solar\footnote{We use solar abundances from \citet{1989GeCoA..53..197A} here also for the intrinsic absorber for a direct comparison with values published in the literature.} metallicity: one Galactic held fixed at $N(\rm{H})_{\rm{X}} = 8.6\times10^{20}\,\rm{cm}^{-2}$ \citep{2005A&A...440..775K}, one intrinsic to the GRB host. The reddening in the optical/NIR wavelength range is modeled by extinction laws from the Magellanic Clouds and the Milky Way in the parametrization of \citet{1992ApJ...395..130P}. 

All synchrotron models yield acceptable fits to the data, and there is mild degeneracy between the parameters, in particular in the case of a broken power-law continuum with a smooth turnover from low to high-energy spectral index. The statistically preferred fit (combined statistics\footnote{We simultaneously use Cash-statistics for the X-ray data and $\chi^2$-statistics for the optical/NIR measurements.} of 302 for 367 d.o.f.) is obtained, however, with a single power-law extending from the NIR to the X-ray band. In the case of a broken power-law continuum the improvement in the fit is small (combined statistics of 299 for 366 d.o.f.), and the best-fit break energy is located at around 3~keV. With respect to the derived value of $A_V$, the single and broken power-law fits are thus equivalent. 

{The SED fit with a single power-law is also consistent with the similar temporal evolution of the optical/NIR and X-ray afterglow light curves (see Figure~\ref{fig:lc}). This provides strong support to the conclusion that both optical/NIR and X-ray data probe a common spectral range of the afterglow's synchrotron radiation.}

The resulting NIR-to-X-ray SED with the single power-law model is shown in Figure~\ref{fig:bbsed}. Given the modest amount of reddening and inferred visual extinction $A_V$, different extinction laws give comparable results. The best-fit is obtained in an SMC parametrization ($\chi^2 = 2.9$ for 5 d.o.f. in the relevant wavelength range), and there is no evidence for the presence of a 2175~\AA~dust feature ($\chi^2 = 7.2$ for LMC, and $\chi^2 = 16.4$ for MW extinction curves). The lack/weakness of a 2175~\AA~dust feature is corroborated by the X-shooter spectrum. While the spectral normalization and large-scale shape is set by our (model-dependent) flux calibration procedure, the inferred slit-loss is uniformly continuous over the full spectral range and does not obliterate spectral features of the size of the 2175~\AA~bump ($\rm{FWHM} \sim 2000$~\AA~observed frame). When fitting the spectrum using an extinction law in the analytical form based on \citet{1990ApJS...72..163F, 2007ApJ...663..320F}, there is again no evidence for a 2175~\AA~dust feature. The $c_3$ parameter of the \citet{1990ApJS...72..163F} extinction curve (representing the strength of the 2175~\AA~bump) is zero within errors. This result is similar to that derived from a sample of GRB afterglows with modest ($\langle A_V\rangle \sim 0.4$~mag) dust obscuration \citep{2012A&A...537A..15S} and consistent with the absence of diffuse interstellar bands (DIBs) in our spectrum \citep[e.g.,][]{2006A&A...447..991C, 2006ApJS..165..138W}.{The $\lambda$6284~DIB is the only transition of the most prominent DIBs at $z=2.36$ that is not located in a window of strong telluric absorption. We set a limit on its equivalent width of $W_{\rm{r}}(\lambda\,6284) <  0.6$~\AA.}

Our data do not constrain the total-to-selective reddening $R_V$, and we report values based on the SMC value ($R_V = 2.93$). Best-fit parameters in the single power-law, SMC case are $A_V= 0.15\pm0.02$~mag, $N(\rm{H})_{\rm{X}}  = (3.5\pm0.2)\times 10^{21}\,\rm{cm}^{-2}$ at $z=2.36$ and a spectral index of $\beta=0.78\pm0.01$, which are very typical values for GRB afterglows in general \citep[e.g.,][]{2006ApJ...641..993K, 2010ApJ...720.1513K, 2011A&A...526A..30G}. Also the ratio between X-ray absorption and SED-derived extinction is fully within the range typically measured for \textit{Swift} GRBs \citep[e.g.,][]{2007MNRAS.377..273S, 2010MNRAS.401.2773S, 2010MNRAS.403.1131N, 2012arXiv1212.4492W}. LMC and MW models yield comparable values  within errors for all parameters.

We caution and stress that these values are strictly upper limits to the host's visual extinction and X-ray absorbing column, because of the presence of strong foreground absorbers (see Section~\ref{intspec}), which contribute to both $A_V$ and $N(\rm{H})_{\rm{X}}$. The expected reddening introduced by the very strong \ion{Mg}{2} absorber at $z = 1.539$, for example, could alone be sufficient to explain the observed reddening in the data \citep[e.g.,][]{2008MNRAS.385.1053M}. We will  thus conservatively consider the derived values for $A_V$ and $N(\rm{H})_{\rm{X}}$ from the broad-band SED as upper limits on the respective parameters in the GRB system in the following.

\subsection{Optical/Near-Infrared Spectrum}
\label{spec}

\begin{figure}
\includegraphics[angle=0, width=0.96\columnwidth]{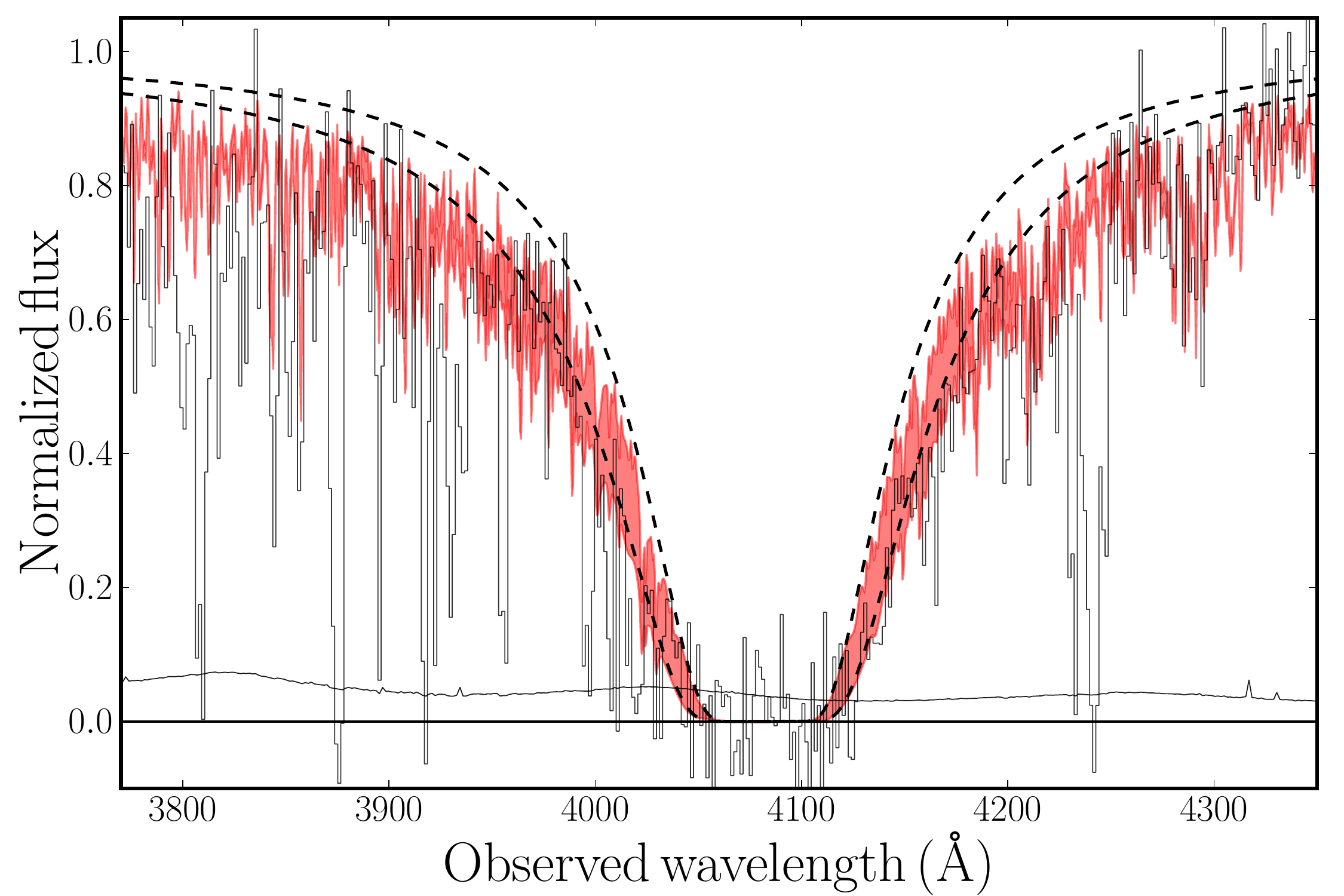}
\caption{X-shooter spectrum around the damped Ly$\alpha$ absorption. Black lines are the X-shooter data and noise level, both binned for clarity. The red-shaded region {shows the fit to the damped Ly$\alpha$ absorption line using the combination of a Voigt profile with $\log(N(\rm{\hi})/\rm{cm^{-2}}) = 21.95\pm0.10$ and the H$_2^{*}$-model from Section~\ref{sec:h2star}. The blue solid and dashed lines illustrate a damped Ly$\alpha$ profile and associated 1$\sigma$ uncertainties, respectively, that correspond to $\log(N(\rm{\hi})/\rm{cm^{-2}}) = 21.95\pm0.10$, but without the vibrationally-excited transitions of molecular hydrogen.}}
\label{fig:dla}
\end{figure}

The X-shooter spectrum of the afterglow of GRB~120815A (Figures~\ref{uvbspec}, \ref{visspec} and \ref{nirspec}) displays a multitude of absorption lines. They are interpreted as Ly$\alpha$, Ly$\beta$, different neutral, low and high-ionization metal lines, several fine-structure transitions, as well as absorption by molecular hydrogen including its vibrationally-excited transitions. The GRB redshift derived from these absorption lines, $z = 2.358$, was previously reported by \citet{2012GCN..13649...1M}. The hydrogen column density is $\log(N(\rm{\hi})/\rm{cm^{-2}}) = 21.95\pm0.10$ (Figure \ref{fig:dla}). Here we used the model of H$_2^*$ (see Section~\ref{sec:h2star}), to estimate its influence on the DLA strength. In addition, emission from the [\ion{O}{3}]($\lambda\,5007$) transition is also detected in the NIR arm of the spectrum (see Section \ref{eml}, and Figure~\ref{fig:eml}). 

\subsubsection{Intervening Systems}
\label{intspec}

We identify five different intervening systems in the X-shooter spectrum that are characterized by at least one transition with $W_{\rm{obs}} > 1$~\AA. In detail, there is a very strong \ion{Mg}{2} absorber at $z = 1.539$ that also shows lines from various other metal transitions (e.g, \ion{Fe}{2}, \ion{Mg}{1}, \ion{Al}{3}). Its rest-frame equivalent width of $W_{\rm{r}}(\lambda\,2796) = 6.1\pm 0.2\,$\AA~makes it the strongest intervening \ion{Mg}{2} absorber yet observed towards a GRB \citep[e.g.,][]{2006MNRAS.372L..38E, 2009A&A...503..771V}. In addition, there are weaker systems at $z = 1.693$ ($W_{\rm{r}}(\lambda\,2796) = 0.50\pm 0.03\,$\AA) and $z = 2.013$ ($W_{\rm{r}}(\lambda\,2796) = 0.48\pm 0.04\,$\AA). Absorption from the \ion{C}{4} doublet is present at $z = 2.000$ ($W_{\rm{r}}(\lambda\,1548) = 0.63\pm 0.04\,$\AA) and $z = 2.338$ ($W_{\rm{r}}(\lambda\,1548) = 0.43\pm 0.04\,$\AA). The intervening systems will not be discussed further in this work. 

\subsubsection{Emission Lines}
\label{eml}

\begin{figure}
\includegraphics[angle=0, width=0.99\columnwidth]{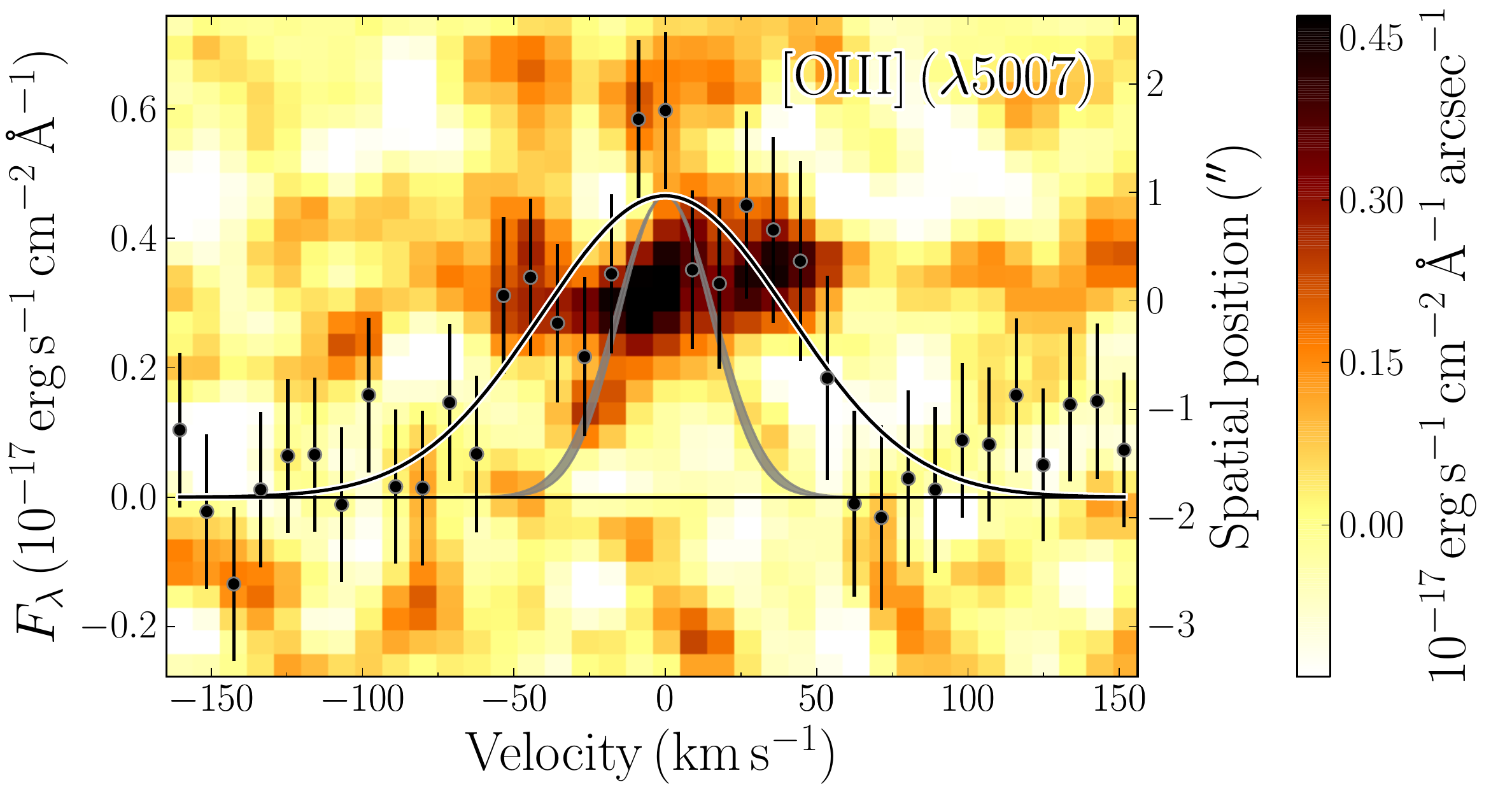}
\caption{One- and two-dimensional, afterglow-subtracted X-shooter spectrum centered around the location of [\ion{O}{3}]($\lambda\,5007$) at $z = 2.3586$ (i.e., observed 16820~\AA). The one-dimensional spectrum (black data) fitted by a Gaussian (black line) is plotted against the {left} y-axis, while the scaling of the two-dimensional spectrum is shown by the color-bar on the right side of the plot. The two-dimensional spectrum has been smoothed by a Gaussian kernel with a FWHM of 1~pixel (0.5~\AA) along the dispersion x-axis, and 2~pixel (i.e., 0\farc{4}) along the spatial y-axis for illustration purposes. The grey-shaded area represents the spectral resolution.}
\label{fig:eml}
\end{figure}

We used spectral point spread function subtraction \citep[e.g.][]{2000Msngr..99...31M, 2010MNRAS.408.2128F} of the afterglow light to search for rest-frame optical emission lines in the X-shooter data. In the afterglow-subtracted spectrum (see Figure~\ref{fig:eml}), we detect [\ion{O}{3}]($\lambda\,5007$) in emission with a significance of $6\sigma$. None of the emission lines of Ly$\alpha$, \ha, \hb,  \oii, \oiii($\lambda\,4959$), \nii~or \ion{He}{2}($\lambda\,4686$) are robustly detected. The Gaussian centroid of [\ion{O}{3}]($\lambda\,5007$) provides a systemic redshift of $z = 2.3586 \pm 0.0001$ for the galaxy hosting GRB~120815A. The line width is relatively well resolved {in velocity space}, and from the observed full-width-half-maximum of $\rm{FWHM} \sim 95\,\rm{km\,s^{-1}}$, we derive a velocity dispersion $\sigma_{\rm vel} = 35_{-9}^{+6}\,\rm{km\,s^{-1}}$ after correcting for the instrumental resolution, and taking into account its uncertainty\footnote{Allowing values between the assumed resolving power and its lower limit derived from arc-lamps, see Section~\ref{xs}.}. 

{Along the spatial axis, the emission from [\ion{O}{3}]($\lambda\,5007$) is unresolved: the FWHM of a Gaussian fit to the [\ion{O}{3}]-line profile is fully consistent with the one of the afterglow continuum emission. Also the respective spatial profiles after dividing the line in a red and blue part are similar within 1.8~standard deviations.} Lacking a measurement of the visible extent of the GRB host, no dynamical mass can be derived here. A comparison to GRB hosts \citep[e.g.,][]{2009ApJ...691..182S, 2012A&A...546A...8K} or lensed galaxies at comparable redshift \citep[e.g.,][]{2011MNRAS.413..643R, 2012arXiv1209.0767C}, however, indicates a dynamical or luminous mass well below $10^{10}\,M_{\sun}$ for the host of GRB~120815A. The total flux (after matching the NIR spectrum to the afterglow model, and including its uncertainty) in the line is $F($\oiii$)=(2.4\pm0.5)\times10^{-17}$~\erg. The 3$\sigma$-confidence limit on the H$\alpha$ flux is $F($H$\alpha) < 4.2\times10^{-17}$~\erg, corresponding to a limit on the observed, i.e., not extinction corrected, star-formation rate $\rm{SFR}<9\,$\Msun$\,\rm{yr}^{-1}$ using the conversion from \citet{1998ARA&A..36..189K} with a \citet{2003PASP..115..763C} initial mass function.

\subsubsection{Absorption Lines, Velocity Structure and Abundances}
\label{sec:absl}

\begin{deluxetable}{cccc} 
\tablecolumns{4} 
\tablewidth{0pc} 
\tablecaption{Column Densities and Abundances\label{tab:lines}}
\tablehead{ \colhead{Ion/molecule}    &    \colhead{Transition$^{\rm{(a)}}$}   & \colhead{Column 
density$^{\rm{(b)}}$}   & \colhead{Abundance$^{\rm{(c,d)}}$} \\ \colhead{} & \colhead{}   & \colhead{$\log(N/\rm{cm^{-2}})\pm \sigma_{\log N}^{(e)}$}
 & \colhead{[X/H]}  }
\startdata
\hi & Ly$\alpha$ & $21.95\pm0.10$ & \nodata  \\ 
H$_2$ & $J = 0$ to $J = 3$ & $20.54\pm0.13$ & \nodata  \\ 
H$_2$ & $J = 0$ & $19.84\pm0.33$ & \nodata  \\ 
H$_2$ & $J = 1$ & $20.43\pm0.12$ & \nodata  \\ 
H$_2$ & $J = 2$ & $16.76\pm0.50$ & \nodata  \\ 
H$_2$ & $J = 3$ & $\lesssim19.01^{\rm{(f)}}$ & \nodata  \\ 
CO & AX$(0-0)$ to $(5-0)$  & $< 15.0$ & \nodata  \\ 
\ion{Zn}{2} &  2026, 2062 & $13.47\pm0.06$ & $-1.15\pm0.12$ \\
\ion{S}{2} & 1250 & $\lesssim16.22\pm0.25^{\rm{(f)}}$ &  $\lesssim-0.89\pm0.26$\\ 
\ion{Si}{2} & 1808 & $\gtrsim16.34\pm0.16^{\rm{(g)}}$ & $\gtrsim-1.16\pm0.19$ \\ 
\ion{Si}{2}$^{*}$ & 1309, 1533 & $14.31\pm0.07$ & \\ 
\ion{Mn}{2i} & 2576, 2594, 2606 & $13.26\pm0.05$ & $-2.15\pm0.12$ \\
\ion{Fe}{2} &  2249, 2260 & $15.29\pm0.05$ &  $-2.15\pm0.12$ \\
\ion{Fe}{2}$^{*}$ &  2333, 2612 & $13.32\pm0.05$ & \nodata  \\
\ion{Fe}{2}$^{**}$ & 2349, 2607 & $13.11\pm0.08$ & \nodata  \\
\ion{Fe}{2}$\,^{4}$F$_{9/2}$ & 2348  & $13.13\pm0.24$ & \nodata  \\
\ion{Ni}{2} &  1370, 1741, 1751 & $14.19\pm0.05$ &  $-1.95\pm0.12$\\
\ion{Ni}{2}$\,^{4}$F$_{9/2}$ &  2217, 2316 & $13.23\pm0.05$ & \nodata \\
\ion{Cr}{2} &  2056, 2062, 2066 &  $13.75\pm0.06$ &  $-1.87\pm0.12$ \\
\ion{Mg}{1} &  2026  &$13.54\pm0.05$ & \nodata \\
\ion{C}{1} &  1560, 1656  &$13.41\pm0.11$ & \nodata \\
\ion{O}{1}$^{*}$ &  1306 & $15.29\pm0.18$ & \nodata \\
\ion{N}{5} &  1238, 1242 & $\lesssim14.80\pm0.21^{\rm{(f)}}$ & \nodata \\
\enddata 
\tablecomments{$^{\rm{(a)}}$ For H$_2$ and CO, this column denotes the different rotational levels of the vibration ground-state  in the case of H$_2$, or the different CO bandheads, respectively.\newline
$^{\rm{(b)}}$ Total column density as the sum of both velocity components (see Section~\ref{sec:absl}, and Figure~\ref{fig:absl}).\newline
$^{\rm{(c)}}$ Assuming a negligible ionization correction, i.e. $N$(Fe) = $N$(\ion{Fe}{2}),  $N$(Zn) = $N$(\ion{Zn}{2}), etc., and using the total column density of hydrogen as $N$(H) = $N(\hi) + 2 N(\rm{H}_2$).\newline
$^{\rm{(d)}}$ Summed over all detected states of the same ion.\newline
$^{\rm{(e)}}$ Quoted uncertainties are the formal errors provided by VPfit in the case of column densities of metals.\newline
$^{\rm{(f)}}$ Considered as upper limit because of blending.\newline
$^{\rm{(g)}}$ Considered as lower limit because of saturation.\newline}
\end{deluxetable} 

\begin{figure}
\begin{center}
\includegraphics[angle=0, width=0.96\columnwidth]{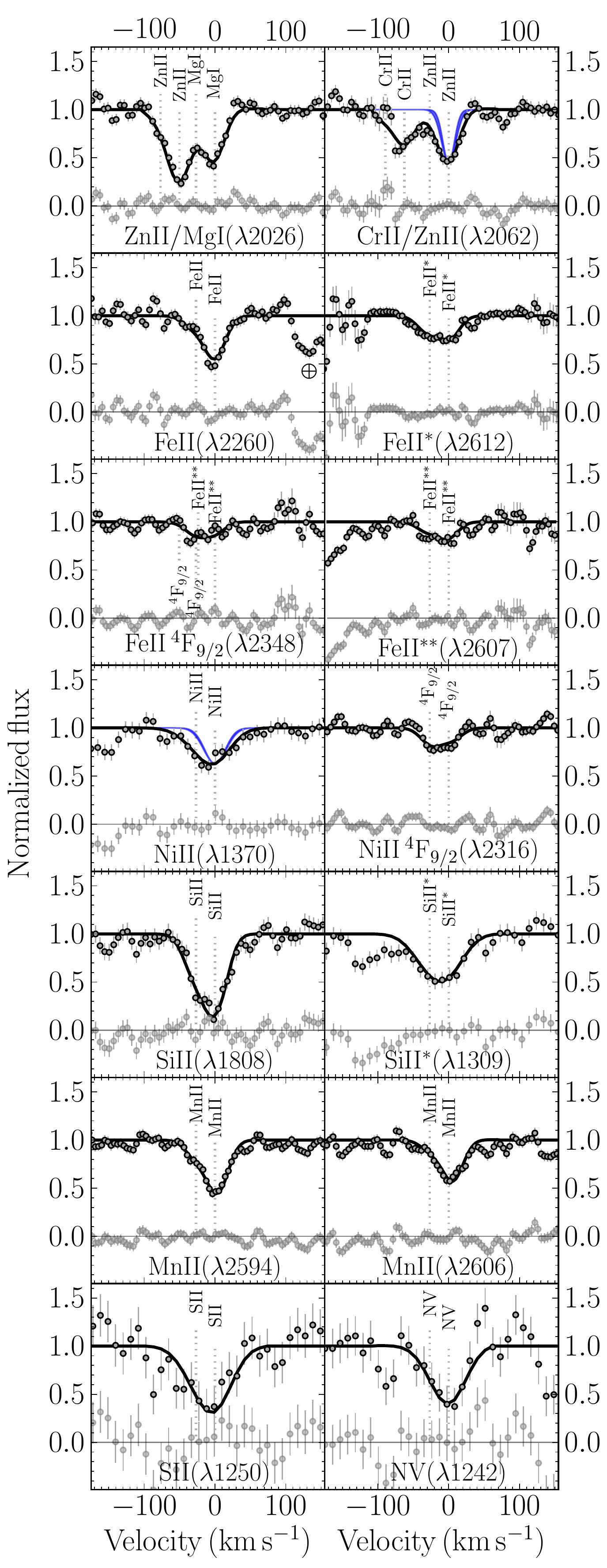}
\caption{Absorption lines and Voigt-profile fits in the GRB~120815A afterglow spectrum. Dark-grey circles and lines show the normalized spectrum with statistical errors and the line fit, respectively. Light-grey circles show the residuals between data and best fit. The instrumental resolution of X-shooter and its error is indicated by shaded, blue areas in panels 2 (VIS arm) and 7 (UVB arm). The position of individual metal lines of the two absorbing systems is indicated by vertical dotted lines. Absorption by strong telluric lines is marked with a circled plus (panel 3).}
\label{fig:absl}
\end{center}
\end{figure}

The absorption lines in the X-shooter afterglow spectrum are relatively narrow, spanning $\Delta V = 83_{-12}^{+20}\,$\kms based on \ion{Si}{2}($\lambda\,1808$) and following the definition in \citet{2006A&A...457...71L}. The absorption is concentrated into two components separated by $\sim 27\, \rm{km\,s^{-1}}$ and, within our resolution and S/N, no further absorption component is required to adequately fit the individual absorption lines (Figure~\ref{fig:absl}). The redshifts of the blue and red component are $z=2.3579\pm 0.0001$ and $2.3582 \pm 0.0001$ respectively, which is 63~\kms and 36~\kms blueshifted with respect to the systemic redshift defined by the \oiii($\lambda\,5007$)~emission line (Section~\ref{eml}).

To derive column densities, we fitted Voigt profiles convolved with the instrumental resolution to unsaturated absorption lines within \texttt{VPfit v.10.0}\footnote{\texttt{http://www.ast.cam.ac.uk/$\sim$rfc/vpfit.html}}. Here, we linked the broadening parameter $b$ between the individual fits of weak absorption lines, and allowed for a small offset ($\lesssim 5-10\,$\kms) in the wavelength calibration between line fits of individual transitions of the same ion. We further assumed that only turbulent broadening contributes to $b$. Finally, we imposed the same velocity structure of two central components for all lines, independent of their excitation level or ionization state. While this is supposedly a simplification of the ISM velocity structure, our data do not yield more detailed constraints (Figure~\ref{fig:absl}).

The broadening parameters are constrained to $b_1=9.7\pm1.0\,$\kms for the primary, red component, and $b_2 = 11.9 \pm 3.5\,$\kms for the weaker, blue component, where the quoted uncertainties on $b$ are the formal errors provided by VPfit. {At the given S/N and instrumental resolution of $\rm{FWHM} = 28$~\kms in X-shooter's VIS arm, features with $b \gtrsim 6\,$\kms (and thus the velocity profiles of the metal absorption) are resolved. The broadening parameters are close to the instrumental resolution in the UVB arm (Figure~\ref{fig:absl})}. The simultaneous fit of all weak lines (excluding the saturated \ion{Si}{2}($\lambda\,1808$)) has a $\chi^2 = 1058$ for {954} degrees of freedom. \ion{Si}{2}($\lambda\,1808$) is fitted after fixing its line properties to the previously-obtained values from weaker lines. 

Given the limited resolution and small separation, the velocity components are blended. The column densities of the single components are correlated and accordingly have substantial errors individually. The total column as the sum of both components, however, is typically well constrained, and we report only total column densities in the following. The column density of metals and its statistical errorbars in Table~\ref{tab:lines} could suffer from systematic uncertainties because of hidden saturation in the medium resolution X-shooter data \citep[see e.g.,][]{2006ApJ...650..272P}. In the absence of data with a very high resolving power, these systematic uncertainties are hard to quantify robustly in our case. There is, however, generally good agreement (within 0.1~dex., except for the saturated \ion{Si}{2}($\lambda\,1808$) transition) between UVES and X-shooter data for the strong ($\log(N(\rm{\hi})/\rm{cm^{-2}}) = 22.10\pm0.10$, $\log ( N ($\ion{Zn}{2}$)/\rm{cm^{-2}} ) \sim 13.6$,  $\log(N($\ion{Fe}{2}$)/\rm{cm^{-2}}) \sim 15.8$) QSO-DLA studied in \citet[][their Table~2]{2012A&A...540A..63N}. \citet{2013arXiv1304.4231K} also find good agreement between QSO-DLA abundances measured in two independent X-shooter spectra of different resolution. Although we cannot rule out that our column densities are affected by hidden saturation, even a factor of four larger error-bars as formally provided by \texttt{VPfit}, or an additional error or underestimation of 0.2~dex. --- the maximum difference in the study of \citet{2012A&A...540A..63N} --- on the abundances of \ion{Si}{2} or \ion{Zn}{2}, do not change the conclusions of this work.

Our best constraint on the metallicity of the GRB-DLA comes from the abundance of zinc, an element generally considered as being  very little affected by dust depletion. Based on the well-detected \ion{Zn}{2}($\lambda\lambda\,2026, 2062$) transitions, we derive $\rm{[Zn/H]} = -1.15\pm0.12$. Estimates from \ion{Si}{2} and \ion{S}{2} agree well within errors. Velocity profiles of selected lines are shown in Figure~\ref{fig:absl} and column densities are given in Table~\ref{tab:lines}. 

All metal transitions bluewards of $\lambda_{\rm{rest}}\sim1700$~\AA~are blended with H$_2^*$ lines (Section~\ref{sec:h2star}). Consequently, we conservatively consider the values derived for weak lines that are in regions of strong H$_2^*$ transitions (in particular \ion{S}{2}($\lambda\,1250$) and to a lesser extent also \ion{N}{5}($\lambda\lambda\,1238,1242$)) as upper limits only. In addition, absorption lines bluewards of $\lambda_{\rm{rest}} < 1260$~\AA~are located in the damping wing of Ly$\alpha$, adding further to the error of the column density because of the uncertainty in their continuum placement. 

\subsubsection{Depletion of Refractory Elements}
\label{sec:depletion}

The measured column densities and abundances (Table~\ref{tab:lines}) of the refractory elements such as Fe, Cr, Ni or Mn with respect to Zn, Si or S indicate that they are depleted onto dust grains. For GRB~120815A, we derive a relative abundance of [Zn/Fe] = $1.01 \pm 0.10$, a column density of iron locked-up in dust of $\log(N\rm{(Fe)^{dust}}/\rm{cm^{-2}}) = 16.27 \pm 0.08$ \citep[following][]{2006A&A...454..151V}, and a dust-to-gas ratio based on zinc $\kappa_{\rm{Zn}} = 10^{\rm{[Zn/H]}} \left( 1 - 10^{\rm{[Fe/Zn]}} \right)$ of $\log(\kappa_{\rm{Zn}}) = -1.19 \pm 0.12$, approximately 40\% of the value of the SMC. The column density of iron in dust grains is a factor of 40 larger than the threshold of $\log(N\rm{(Fe)^{dust}}/\rm{cm^{-2}}) > 14.7$ above which H$_2$-molecules are typically found in QSO-DLAs \citep[e.g.,][]{2008A&A...481..327N}.

Using the SMC as a benchmark\footnote{This requires assumptions on the SMC's dust depletion ($\rm{[Fe/Zn]}_{\rm{SMC}} = -1.0$), and metallicity ($\rm{[M/H]}_{\rm{SMC}} = -0.7$), the average reddening per $N(\hi)$ column, and the total-to-selective reddening, which were all taken from \citet{2007ApJ...666..267P} and references therein.}, the depletion-derived extinction is $A_V = 0.12 \pm 0.03$~mag. The metals-to-dust ratio $(\log(N(\hi)/\rm{cm}^{-2})+\rm{[Zn/H]})/A_V = (5\pm1)\times10^{21}\,\rm{cm^{-2}\,mag{^{-1}}}$ is roughly 2.5 times the value of the Local. Different depletion methods \citep[see, e.g.,][]{2003ApJ...585..638S} and comparison values from the Local Group allow visual extinctions up to $A_V\sim0.6$~mag, and consequently a metals-to-dust ratio based on the depletion analysis that is consistent with the Galactic value \citep[see e.g,][]{1978ApJ...224..132B, 1995A&A...293..889P}.

\subsubsection{Molecular Hydrogen}
\label{sec:molhydro}

\begin{figure*}
\includegraphics[angle=0, width=1.95\columnwidth]{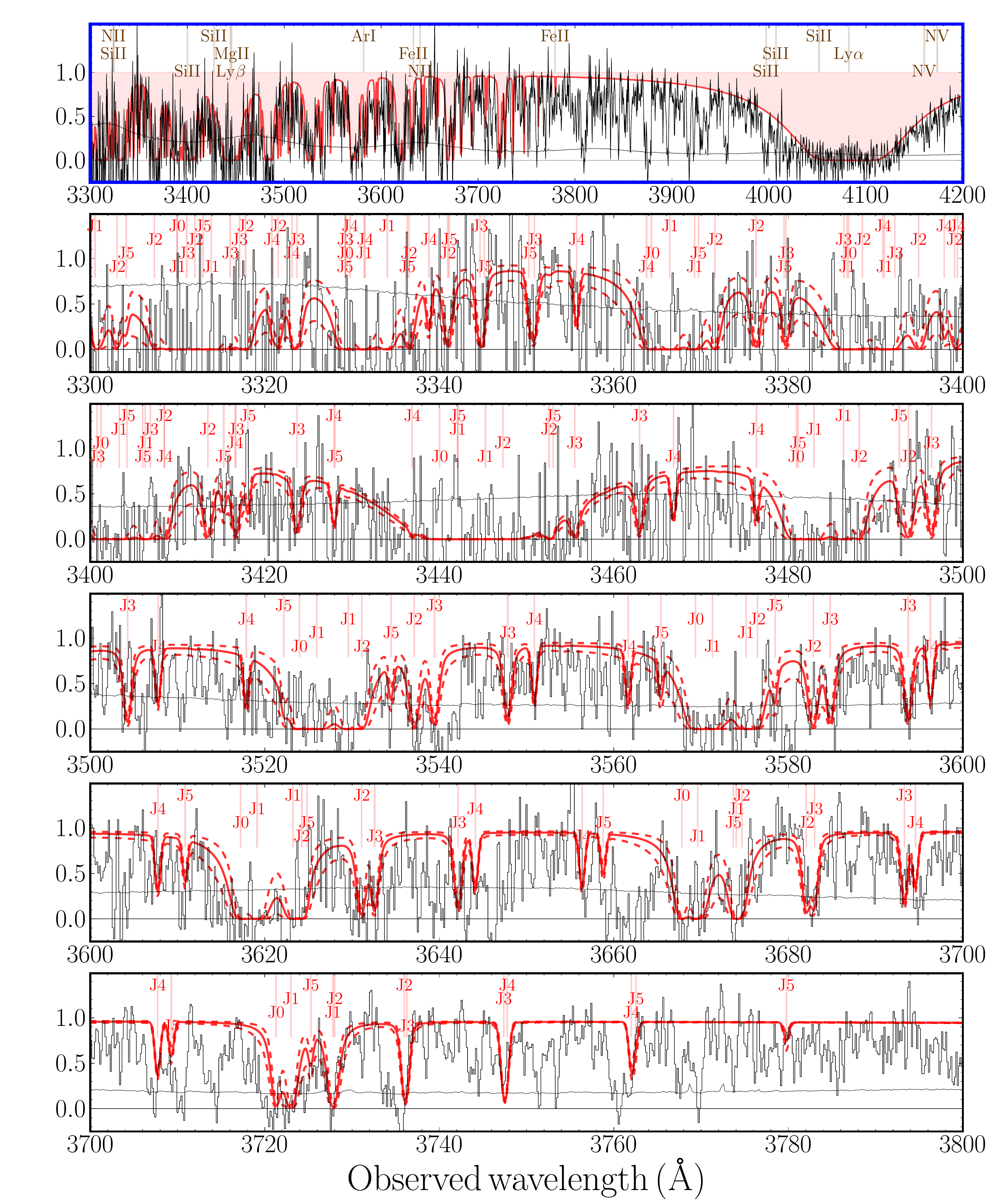}
\caption{X-shooter spectrum between 3300 and 4200~\AA~illustrating the presence of H$_2$ absorption. The uppermost, blue-framed panel shows an overview, while the lower 5 panels show zoom-ins of 100~\AA~each below 3800~\AA, where most of the Lyman-Werner absorption bands are located. Light-grey lines are the normalized spectrum, while dark-grey lines indicate the error spectrum. The solid red line denote the synthetic H$_2$ model, with individual transitions up to $J=5$ transitions marked with red labels. {The dashed lines show synthetic H$_2$ models corresponding to the 1$\sigma$ errors on the measured molecular content.} In the top panel, we also mark prominent metal absorption lines previously detected in GRB-DLAs \citep{2011ApJ...727...73C} and those of the intervening systems. In the lower panels, J0, J1, and so forth denote transitions from the $J=0$, $J=1$ rotational level of the vibrational ground state of the Lyman-Werner bands of H$_2$.}
\label{fig:h2mod}
\end{figure*}

The most salient feature of the X-shooter spectrum blueward the Ly$\alpha$ absorption is the presence of periodic regions of negligible flux (see Figure~\ref{fig:h2mod}, top panel). They are interpreted and well described by the Lyman-Werner absorption bands of molecular hydrogen, rotationally-excited states between the Lyman-limit and a rest-frame wavelength of around $1110~$\AA.

The fitting and analysis of the molecular hydrogen transitions follows \citet{2002A&A...392..781L, 2003MNRAS.346..209L}. Given the available data, we can constrain the column densities of the rotational levels $J=0$ up to $J=3$ in the first five Lyman bands, $L=0-0$ to $0-4$. {We first performed a Voigt-profile fit of the $J=0$, 1, 2 and 3 lines independently. The resulting column density of the $J=3$ level is considered as an upper limit due to possible blending}. In total, these levels yield an integrated H$_2$ column density of $\log(N(\rm{H_2})/cm^{-2}) = 20.54 \pm 0.13$. Details of individual rotational transitions are provided in Table~\ref{tab:lines}. {Higher $J$-levels, which are expected to be populated by fluorescence, and their relation to the excited H$_2^*$ (Section~\ref{sec:h2star}) remain unconstrained with the available X-shooter data because of the medium resolution, limited S/N, and blending with Ly$\alpha$ forest lines. Our data provide a robust measurement only of the excitation temperature of the core of the cloud (which is therefore unrelated to fluorescence).} The kinetic temperature for the excitation of the $J=1$ state is $T_{\rm kin} \sim 200\,\rm{K}$, which is higher than observed locally \citep[e.g.,][]{1977ApJ...216..291S, 2002ApJ...566..857T}, but in the range of QSO-derived temperatures \citep[][]{2005MNRAS.362..549S}. 

Together with the column density of neutral hydrogen $\log(N\rm{_H/cm^{-2}}) = 21.95 \pm 0.10$, this translates into a molecular fraction of $\log f = -1.14 \pm 0.15$, where $f \equiv 2N(\rm{H}_2)$/$(N(\hi)+ 2N(\rm{H}_2))$. 
The best-fit redshift and broadening parameter of the molecular hydrogen lines are $z(\rm{H}_2) = 2.3582$ and $b({\rm{H}_2}) = 8.7 \pm 0.6\,$\kms, consistent with the primary component of the metal-line absorption (Section \ref{sec:absl}). {Using the measured parameters and errors on the kinetic temperature, broadening parameter and molecular fraction we then calculated synthetic models that include higher rotational levels and are shown in Figure~\ref{fig:h2mod}, lower panels.}

\subsubsection{Lack of Carbon Monoxide}
\label{sec:co}

We do not detect absorption signatures from the CO molecule. While there is an absorption line at $\sim4960$~\AA~with observed equivalent width $W_{\rm obs} = 0.64 \pm 0.10$~\AA, conspicuously close to the CO AX(2-0) bandhead at $z=2.36$, we consider it unrelated to CO, as none of the other expected CO transitions are detected. Also, the CO redshift would be significantly offset from the value of either absorption, emission or H$_2$. We thus conclude that this single line is not caused by CO, but related to an intervening absorber. Given the presence of a second absorption line at $\sim4966$~\AA~with $W_{\rm obs}(\lambda\,1550) = 0.33 \pm 0.07$, the most likely interpretation of these features is the \ion{C}{4} doublet at $z = 2.2042$. 

Using the wavelength range of the six strongest CO AX bandheads (CO AX($0-0$) to CO AX($5-0$)) redshifted to $z = 2.3582$ as reference, we set a $3\sigma$ upper limit on the total CO column density\footnote{Using molecular data for CO from \citet{1994ApJS...95..301M}.} of $\log(N(\rm{CO)/cm^{-2}}) < 15.0$. This value is fairly independent on the CO excitation temperature (assumed to be greater than $T_{\rm{CMB}}=9\,$K at $z=2.36$) and CO broadening parameter, because the latter would not be resolved for typical values with our X-shooter data \citep[e.g.,][]{2010A&A...523A..80N}. 

The $N$(CO) to $N$(H$_2$) ratio is smaller than $10^{-5.5}$, which is typical for Galactic measurements with $f < 0.1$ \citep{2010ApJ...708..334B}. The H$_2$-bearing system towards GRB~120815A is thus characterized as a diffuse cloud, consistent with the small amount of reddening and extinction (in comparison to Galactic sightlines; \citealp{2006ARA&A..44..367S}).

{$N$(CO)/$N$(H$_2$) is a strong function of the environment,} and because of the sub-solar GRB-DLA metallicity, could even be below what is typically observed in the Milky Way. {The CO luminosity per unit luminosity or per unit SFR of local dwarf galaxies, for example, is significantly lower than extrapolated from metal-rich spirals \citep{2012AJ....143..138S}. This can be interpreted as evidence for a strong metallicity dependence of the CO to H$_2$ ratio below half of $Z_{\sun}$ \citep[see also, e.g.,][]{2011ApJ...737...12L}. Measurements from GRB-DLAs similar to the one presented in this work {and in \citet{2009ApJ...691L..27P}} will help to probe and study the CO abundance and the H$_2$ to CO ratio in {very different environments from diffuse clouds to translucent sight-lines} over a broad range of metallicities (or stellar masses) and redshifts directly and complementary to conventional studies \citep[e.g.,][]{2012ApJ...746...69G}.}

Our upper limit on $N$(CO)/$N$(H$_2$) is comparable to or lower than derived from CO detections in high-redshift systems \citep{2008A&A...482L..39S, 2009ApJ...691L..27P, 2010A&A...523A..80N}, which we readily relate to the lower metallicity and dust-extinction of the GRB~120815A DLA as compared to the sightlines in which CO was seen. In QSO-DLAs without CO detection, the constraints on CO/H$_2$ extend down to limits of $N$(CO)/$N($H$_2)<10^{-8}$ \citep[e.g.,][]{2002MNRAS.332..383P}. 

\subsubsection{Detection of Vibrationally-excited H$_2$}
\label{sec:h2star}

\begin{figure}
\includegraphics[angle=0, width=0.99\columnwidth]{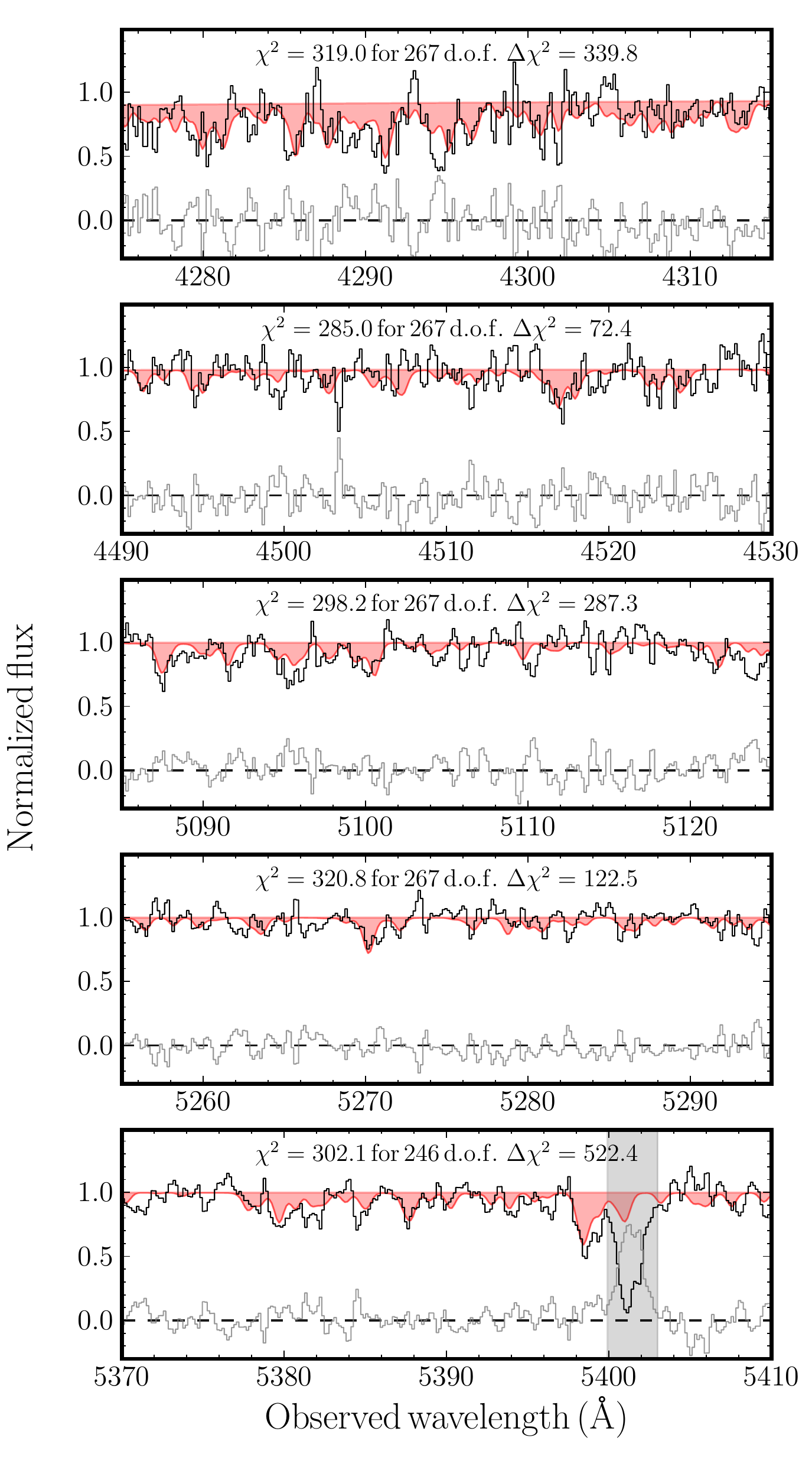}
\caption{{Several cut-outs of illustrative parts of the normalized X-shooter UVB arm spectrum showing the presence of H$_2^*$ lines. The regions have been chosen to have at least one strong transition of H$_2^*$ and to be little affected by blending. The black lines show the data, red lines the best-fit model, and grey lines the residuals. Grey shaded regions are ignored in the fitting because of blending with metal absorption lines. In each panel we also give the best-fit $\chi^2$, as well as the improvement $\Delta \chi^2$ as compared to an unmodified spectrum.}}
\label{fig:h2det}
\end{figure}

\begin{figure}
\includegraphics[angle=0, width=0.99\columnwidth]{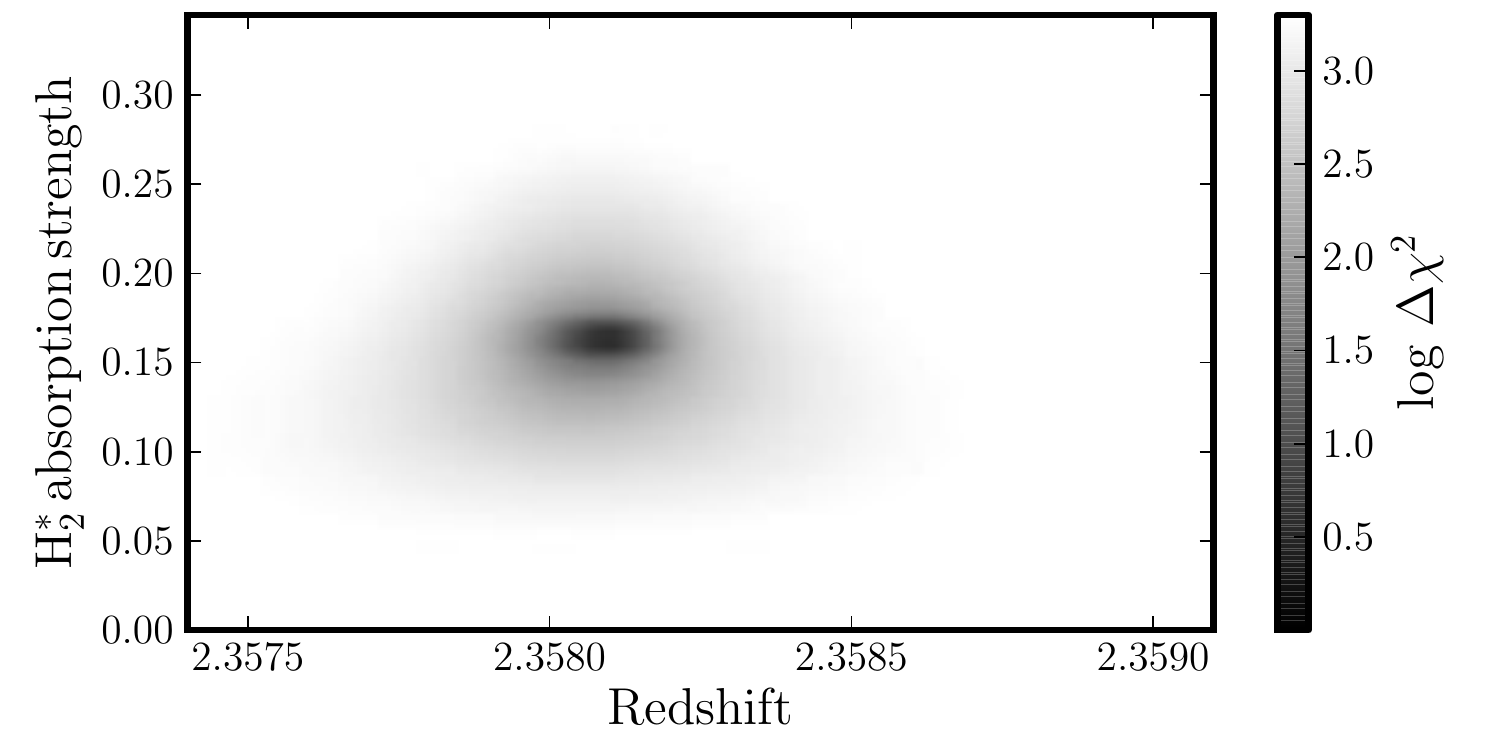}
\caption{Logarithmic $\Delta \chi^2$ distribution for different H$_2^*$ model redshifts and H$_2^*$ absorption line strengths (the ratio between the applied optical depth and the one in the input model).}
\label{fig:h2chi2}
\end{figure}

After establishing the presence of molecular hydrogen, we search for vibrationally-excited H$_2$ by using the synthetic spectrum\footnote{\texttt{http://www.astro.princeton.edu/$\sim$draine/grbh2.html}} calculated in \citet{2000ApJ...532..273D} and \citet{2002ApJ...569..780D}. We downgraded a theoretical $R=10\,000$ spectrum to the resolution of X-shooter's UVB arm ($R\sim6000$), limited the wavelength range to regions above the red damping wing of Ly$\alpha$, and excluded strong metal absorption lines caused by the GRB-DLA and the intervening systems in the further analysis. 

Cross-matching the downgraded spectrum with our data (see Figure~\ref{fig:h2det} for illustrative examples) using different optical depths $\tau$ with respect to the input model and varying redshifts, we find a single and well-defined minimum in the $\chi^2$-space at a redshift consistent with the ground-state H$_2$ and metal transitions (Figure~\ref{fig:h2chi2}). The best match is obtained at $z=2.3581\pm0.0001$, and the presence of H$_2^*$ as compared to an unmodified spectrum is unambiguous (Figures~\ref{fig:h2det} and \ref{fig:h2chi2}). {Because the H$_2^*$-states are overlapping in large parts of the spectrum bluewards of 5400~\AA , we caution that there is a mild degeneracy between the continuum level and the strength of the H$_2^*$-absorption. An overestimation of the continuum would lead to a lower value for the significance of the H$_2^*$ detection as well as for the optical depth $\tau$. Given the presence of several strong individual lines as illustrated in Figure~\ref{fig:h2det}, we consider the detection of H$_2^*$ despite the uncertainty in the continuum placement highly significant.} 

While the overall match between the input spectrum and data is reasonable, there are subtle differences between the data and synthetic model. These are evident in the different line ratios between vibrational levels (see, e.g., the expected but undetected strong line at $\sim4560$~\AA~in Figure~\ref{fig:h2*mod}). This is likely related to different initial conditions --- such as particle densities, cloud-afterglow distance (see Section~\ref{sec:distance}), dust shielding, afterglow luminosity and spectrum, and/or background UV-radiation field --- with respect to the assumptions in the theoretical calculation or uncertain transition probabilities between different excited levels of the H$_2$ molecule. In particular, we derive a distance between the absorbing gas cloud and GRB~120815A of $d \sim  0.5$~kpc (see Section \ref{sec:distance}), much larger than the $d \sim 1$~pc assumed in the model. This will inevitably lead to different line strengths and ratios between the various H$_2^{*}$ transitions. {A more detailed analysis and modeling of the H$_2^{*}$ absorption lines will be presented in a forthcoming work.}

\section{Discussion}
\subsection{Distance between GRB and Absorbing Cloud}
\label{sec:distance}

Our observations indicate that the metal-lines and H$_2$ absorption are not caused by circumburst material. The presence of H$_2$ alone sets a lower limit on the burst-cloud distance of at least $d \gtrsim 10$~pc \citep{2008ApJ...682.1114W, 2009A&A...506..661L}. In addition, the detection of neutral species at the same redshift, in particular \ion{Mg}{1} (Figure~\ref{fig:absl}), is generally interpreted as an indication that the gas is at substantial distances ($d>50\,\rm{pc}$) from the GRB \citep{2006ApJ...648...95P}.

{These initial considerations are confirmed by modeling the photo-excitation of the \ion{Fe}{2}$^*$, \ion{Fe}{2}$^{**}$, \ion{Fe}{2}$\,^4\rm{F}_{9/2}$ and \ion{Ni}{2}$\,^4\rm{F}_{9/2}$ transitions. {In the modeling, we excluded \ion{Si}{2}$^*$, because excited states of \ion{Si}{2} are also seen in very actively star-forming galaxies \citep[e.g.,][]{2002ApJ...569..742P}. \ion{Si}{2} is easier to excite than \ion{Fe}{2} or \ion{Ni}{2}, and a fraction of the \ion{Si}{2} excitation might thus not be due to the GRB afterglow radiation. The observed ratio of $N$(\ion{Si}{2}$^*$)/$N$(\ion{Si}{2}) $\sim$ 1\% could, for example, be provided by electron densities around 10~$\rm{cm}^{-3}$ in a $T\sim10^{4}$~K medium, or UV-backgrounds about 10$^{3}$ times that of the Galaxy \citep{2002MNRAS.329..135S}.}. Given the small separation in velocity of both absorption components, and the limited resolution of our X-shooter data, we only model the total column densities \citep[see also the discussion in][]{ 2013A&A...549A..22V}.}

{The photo-excitation modeling follows closely the methodology presented and applied in \citet{2007A&A...468...83V, 2013A&A...549A..22V} and \citet{2009A&A...506..661L}, and we refer to these publications for a detailed description of the technique. As an input to the model, we use the optical afterglow light curve (see Section~\ref{grond}), the afterglow spectral slope $\beta$, the inferred extinction $A_V$ (Sections~\ref{agmodel} and \ref{sec:depletion}) and the broadening parameter $b$ of the absorption lines (\ref{sec:absl}). The dust is placed inside the absorbing cloud that is being excited, and the distance $d$ from the GRB is defined to the near-side of the cloud.}

A good fit ($\chi^2/$d.o.f = 2.3/4) is obtained for a distance between gas cloud and burst of $d = 460 \pm 60$~pc in the case of atomic data from \texttt{CLOUDY} \citep{1999ApJS..120..101V}. A similar result is obtained ($\chi^2/$d.o.f = 4.3/4, $d = 530 \pm 80$~pc) when using atomic data as described in \citet{2007A&A...468...83V}. For a detailed comparison between the different sets of atomic data in this context, we refer to the discussion in \citet{2009A&A...506..661L}. Given the lack of a spectral time series, and relatively few measurements of excited metal transitions, the cloud size is unconstrained by our data. The neutral gas seen in absorption in the GRB-DLA is thus not directly related to the GRB progenitor's immediate environment or stellar wind-driven mass loss.

\subsection{Detection of High-Ionization Species}
\label{sec:nv}

We also detect the imprint of the highly-ionized transition \ion{N}{5} (ionization potential for creation of 77.5~eV) with a high column density at a redshift consistent with the other metal-lines. \ion{N}{5} is a common feature observed towards GRBs \citep{2008A&A...491..189F} and is suggestive of a circumburst origin in the vicinity of the GRB \citep{2008ApJ...685..344P}. \ion{N}{5} is, however, also present in UV spectra of O-type stars, galaxies and QSO-DLAs \citep[e.g.,][]{2002ApJ...569..742P, 2009A&A...503..731F}. In the case of GRB~120815A, in particular, the \ion{N}{5} measurement is, due to blending with H$_2^*$, strictly an upper limit, suffers from low S/N and has thus large errors on its column density as well as uncertainties in its velocity structure. Our detection of \ion{N}{5} is broadly consistent with the measurements for other GRB-DLAs, and does not necessarily point to an origin in the circumburst environment, but could also be of interstellar origin as argued in \citet{2008A&A...491..189F}. A location in the ISM is supported by the non-variability of \ion{N}{5} in a spectral time-series for GRB~080310 \citep{2012A&A...545A..64D}. 

\subsection{Molecular Hydrogen in GRB-DLAs}
\subsubsection{Metallicity and Neutral Hydrogen Column Density}
\label{sec:h2nh}

\begin{figure}
\includegraphics[angle=0, width=0.96\columnwidth]{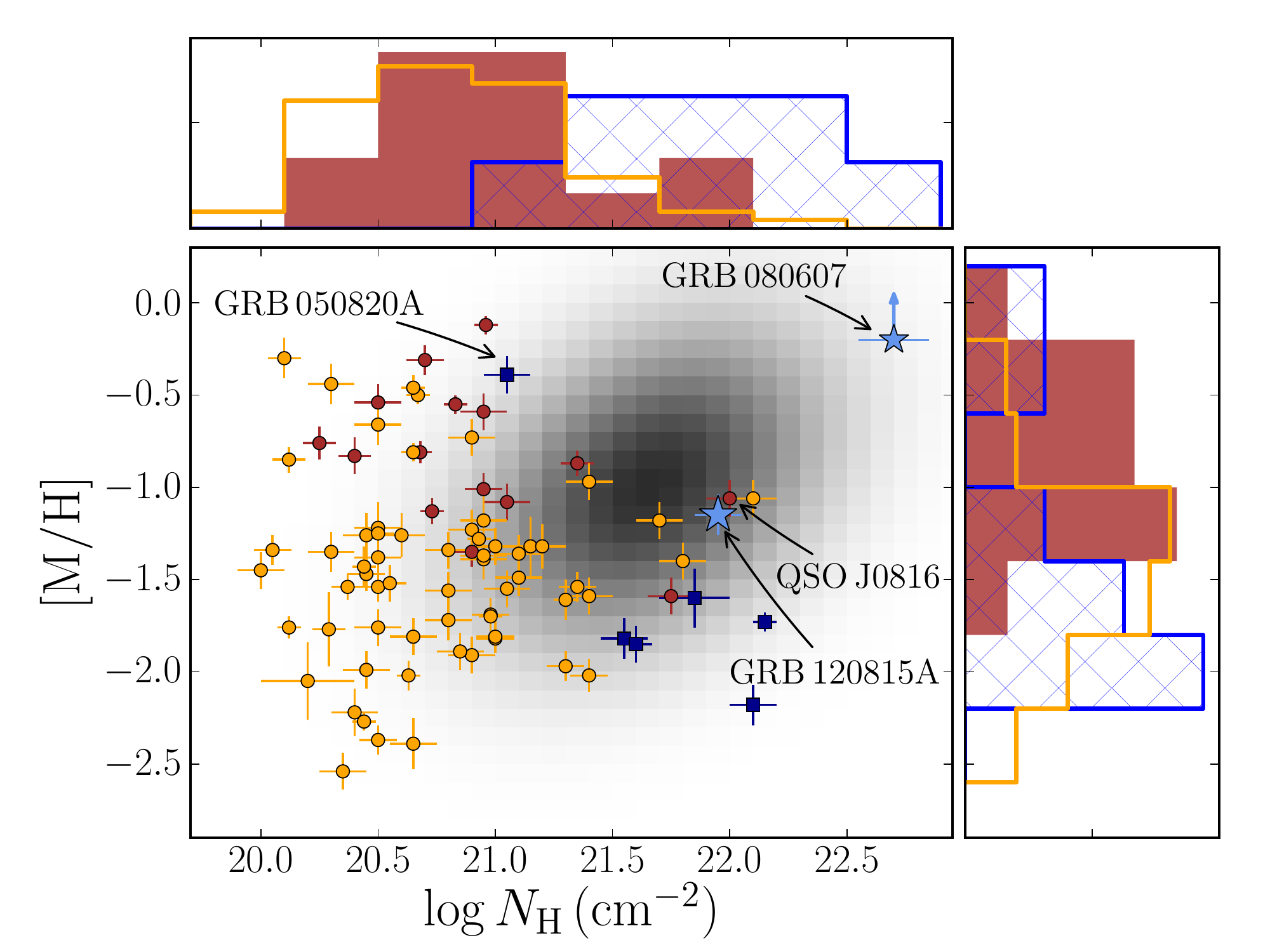}
\caption{The central panel shows metallicity versus hydrogen column density for QSO and GRB DLAs in which H$_2$ was extensively searched for. Circles are QSO-DLAs, where red and yellow circles correspond to DLAs with and without significant molecular hydrogen. Blue data are GRB-DLAs and the blue stars are  GRB~120815A and GRB~080607, where H$_2$ has been detected. The grey background shading indicates the parameter space where many of the more dust-extinguished GRB-DLAs are located {(see details in Section~\ref{sec:h2grb})}. The upper and right panels illustrate the respective normalized histograms. Blue hatched histogram are GRBs including both H$_2$-poor and H$_2$-rich sightlines due to low number statistics. Red-filled and yellow histograms are QSO-DLAs with and without molecular hydrogen, respectively. Data for QSOs are from \citet{2008A&A...481..327N, 2010MNRAS.408.2128F, 2011MNRAS.413.2481F, 2012AJ....143..147G}, data from GRBs from \citet{2009A&A...506..661L, 2009ApJ...691L..27P, 2010A&A...523A..36D}.}
\label{fig:dlasnhmet}
\end{figure}

The column densities of $\hi$, H$_2$ and the molecular fraction $f$ of the DLA towards GRB~120815A are high when compared to the average properties\footnote{A few noteworthy examples of dusty and metal-enriched  QSO-DLAs \citep[e.g.,][]{2011MNRAS.413.2481F} or QSO-DLAs with properties similar to those of typical GRB-DLAs  \citep[e.g.,][]{2012AJ....143..147G} exist.} of the sample of QSO-DLAs. {In the work of \citet{2012A&A...547L...1N}, for example, there are 10 out of $\approx 12\,000$ SDSS-selected QSO-DLAs with  $\log(N(\hi)/\rm{cm}^{-2}) \gtrsim 21.95$. The high $N(\hi)$ is, however, more typical for GRB-DLAs (8 of 27 GRB-DLAs have $\log(N(\hi)/\rm{cm}^{-2}) \gtrsim 21.95$ in the sample of \citet{2009ApJS..185..526F}).}

When considering the significant dust depletion and metal abundance along the GRB~120815A sightline, a consistent picture is emerging: High molecular fractions ($\log(f) > -3$) are exclusively seen at significant metallicities ($\rm{[M/H]} \gtrsim -1.5$), while they are not restricted to the very high-end of hydrogen column densities of $\log(N(\hi)/\rm{cm}^{-2}) \sim 21.0$ \citep[see, e.g.,][]{2003MNRAS.346..209L}. 

The high column densities of neutral hydrogen typical for GRB-DLAs are thus not the primary condition for the presence of H$_2$ in DLAs. This is illustrated in Figure~\ref{fig:dlasnhmet}, which shows the incidence of H$_2$-bearing gas as functions of $\hi$ column densities and metallicity. While there is an obvious discrepancy in the distributions of $N(\hi)$ between GRBs and QSO-DLAs, the difference between H$_2$-poor and H$_2$-rich QSO sightlines with respect to their $N(\hi)$ is much more subtle (Figure~\ref{fig:dlasnhmet}, top panel). 

\begin{figure}
\includegraphics[angle=0, width=0.96\columnwidth]{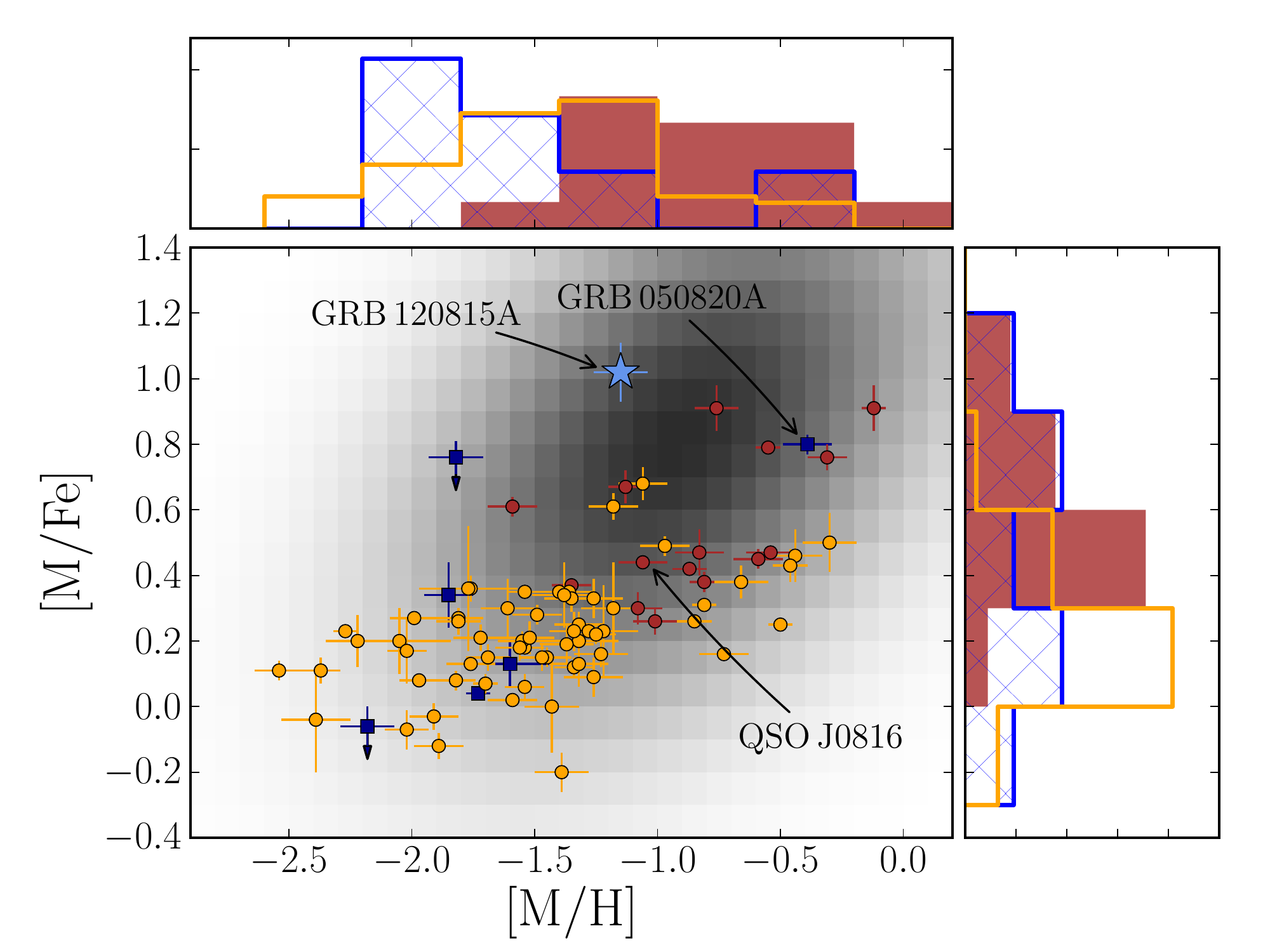}
\caption{Depletion of QSO and GRB DLAs versus their metallicity. Symbols, colors and shadings are the same as in Figure~\ref{fig:dlasnhmet}. GRB~080607 is not shown in this plot because no constraints on the depletion factor are available \citep{2009ApJ...691L..27P}.}
\label{fig:dlasmetdep}
\end{figure}

\subsubsection{The Role of Dust}
\label{sec:h2av}

Dust plays a major role for the presence of molecular gas by catalyzing the formation of molecules on the surface of dust grains as well as shielding against Lyman-Werner photons. For GRB~120815A, the presence of dust is evident from the relative abundance of Fe, Ni, Mn and Cr with respect to Zn ($\rm{[Zn/Fe]} = 1.01 \pm 0.10$). This value is larger than the relative abundance of previous GRB-DLAs observed at high spectral resolution, for which elaborate H$_2$-searches were performed \citep{2009A&A...506..661L}.

Similar to the case of metallicity, there is a higher probability of finding molecular hydrogen for more dust-depleted QSO sightlines (Figure~\ref{fig:dlasmetdep}). There seems to be a lower limit of [M/Fe] of $\sim0.3$ below which no strong signatures of H$_2$-molecules (neither towards QSO nor GRBs) are observed (Figure \ref{fig:dlasmetdep}). The values of metallicity, dust depletion and $N(\hi)$ are effectively combined into a column density of iron locked up in dust $N\rm{(Fe)}^{\rm {dust}}$, which has been identified as a primary driver of H$_2$ detection in QSO-DLAs \citep{2008A&A...481..327N}.

\citet{2012AJ....143..147G} studied the QSO-selected system {SDSS J081634+144612} and identified it as an analog to GRB-DLAs with very similar $N(\hi)$ and [Zn/H] as measured for the DLA of GRB~120815A (Figure~\ref{fig:dlasnhmet}). The column density of $N$(H$_2$) in the QSO-DLA, and thus the molecular fraction $f$, however, is two orders of magnitude lower than derived for GRB~120815A. This could possibly be related to the lower depletion. The mean relative abundance\footnote{Converted into our metallicity scale based on \citet{2009ARA&A..47..481A}.} in {this} QSO-DLA was $[\mathrm{Zn/Fe}] = 0.48 \pm 0.06$, while we observe $[\mathrm{Zn/Fe}] = 1.01 \pm 0.10$ for GRB~120815A (Figure~\ref{fig:dlasmetdep}).

\subsubsection{Implications for the Population of GRB-DLAs}
\label{sec:h2grb}

The region of elevated metallicity, depletion and $N(\hi)$ is exactly the parameter space where we expect many of the more dust-extinguished, and thus UV-fainter afterglows to be located. This combination directly leads to a high absolute $N\rm{(Fe)^{dust}}$, and thus an efficient conversion from $\hi$ to H$_2$ on the dust grain surface \citep{2001ApJ...562L..95S, 2004ApJ...611...40C}. A high column density of dust absorbs UV-photons efficiently, and is also a  necessary condition for creating molecular hydrogen on dust grains. The observed H$_2$ column density is {also affected} by photo-dissociation, for example by the background UV-radiation field in the GRB host. A lack of molecular gas in GRB-DLAs with respect to QSO-DLAs of similar $N\rm{(Fe)^{dust}}$ would thus indicate that photo-dissociation of H$_2$ is more prominent in GRB hosts than it is in the galaxy counterparts of GRB-DLAs. In such a scenario, H$_2$ needs to be absent already prior to the burst, because the derived distances between the absorbing gas and GRB are much larger than the region of $d \lesssim 10\,$pc, in which the afterglow's UV-flux photo-dissociates H$_2$. The previous lack of molecular gas in GRB-DLAs, however, is consistent with their observed metallicity and dust depletion \citep{2009A&A...506..661L}, and does not necessarily point to a difference between GRB and QSO-DLAs with respect to the presence of H$_2$. In fact, GRB~050820A is still the only H$_2$-less GRB-DLA with physical properties that are similar to those of H$_2$-bearing QSO-selected systems (Figure~\ref{fig:dlasmetdep}).

The measured hydrogen column density, metallicity and dust-depletion of GRB~120815A are fairly representative of GRB-DLAs previously observed at lower spectral resolution. {Based on the statistical samples} compiled for $\log(N(\hi)$ in \citet{2009ApJS..185..526F, 2011ApJ...727...73C}, metallicity in \citet{2010ApJ...720..862R, 2011ApJ...727...73C}, and depletion in \citet{2003ApJ...585..638S, 2006NJPh....8..195S}, we expect the distributions of the respective physical quantity to be centered roughly\footnote{We note that also in the low-resolution sample, biases towards lower metallicity and lower depletion values are present \citep{2009ApJS..185..526F}, and that a standard analysis of low-resolution spectra tends to underestimate the true column densities \citep{2006ApJ...650..272P}. Both effects would shift the distribution of metal column densities to even higher values, further strengthening the result.} {around $\log(N(\hi)/\rm{cm}^{-2}) \sim 21.6\pm0.6$,  $\rm{[M/H]}\sim -1.0 \pm 0.8$ and $\rm{[M/Fe]}\sim 0.8 \pm 0.7$, where the quoted error indicates an estimate of the dispersion of the respective sample distribution}. This parameter space is shown by the grey shaded area in Figures~\ref{fig:dlasnhmet} and \ref{fig:dlasmetdep}. {QSO-DLAs in this region show a higher  probability of having significant column densities of molecular hydrogen.} The molecular content of previous GRB-DLAs, however, could typically not be probed with afterglow spectra of low S/N, low resolution, or both. Once we have gained better observational access to afterglows {in this region} through detailed observations with sensitive spectrographs of high-enough resolution (such as X-shooter or UVES), it seems likely that a higher fraction of those GRBs will also show the presence of molecular gas.

\subsection{Vibrationally-excited H$_2$ in GRB-DLAs}
\label{sec:h2*grb}

The presence of vibrationally-excited states of molecular hydrogen (H$_2^*$) along GRB sightlines was first postulated by \citet{2000ApJ...532..273D} and detailed in \citet{2002ApJ...569..780D}. The strong features, however, that were expected to be produced by the vibrationally-excited H$_2$ from the GRB's birth cloud, and detectable in even low-resolution spectra, were never clearly observed. Excited states would nevertheless be present in cases where the molecular cloud only intersects the GRB sightline \citep{2002ApJ...569..780D} at small enough distances. Indeed, while in the work of \citet{2002ApJ...569..780D} the total H$_2^*$ column density produced through UV-pumping is $N\rm{(H_2^*)} = 10^{19}\,\rm{cm}^{-2}$, the best overall match is obtained here with a factor 6 decreased optical depth of the H$_2^*$ transitions compared to the input model. In the case of GRB~080607, \citet{2009ApJ...701L..63S} perform a dynamic modeling of the H$_2^*$ absorption, and derive a distance between GRB~080607 and its H$_2$-bearing cloud between 230 and 940 pc. This is comparable to our estimate of the distance between GRB~120815A and its DLA of $0.5\pm0.1$~kpc.

As shown in Figures~\ref{fig:dla} and \ref{fig:h2*mod}, the H$_2^*$ transitions affect the measurement of $N(\hi)$. If H$_2$ and H$_2^*$ are more common in GRB-DLAs as speculated above, $N(\hi)$ could have been overestimated in previous cases. In the case of GRB~120815A, the H$_2^*$ transitions decrease the best-fit column density by $\sim$0.15~dex. Similarly, blending with H$_2^*$ could be a serious concern for intrinsically weak metal lines, such as \ion{S}{2}($\lambda\lambda\,1250, 1253$) \citep[see also][]{2009ApJ...701L..63S}.

H$_2^*$ opens a route to establish a positive presence of molecular hydrogen also in those cases where the observations do not cover the wavelength range bluewards of Ly$\alpha$, and thus particularly in afterglow spectra of GRBs in the redshift range $1 < z \lesssim 2$ . Strong vibrationally-excited levels of H$_2$ extend up to $\lambda_{\rm rest}\sim 1600\,$\AA, and in contrast to the Lyman-Werner bands at $\lambda_{\rm rest}\lesssim1120\,$\AA, their identification is not compromised by the Ly$\alpha$ forest, and could be performed also with lower-resolution data. Primary candidates for further H$_2^*$ (and thus H$_2$) searches are the high-quality, high-resolution spectra of low-redshift afterglows (in which the individual lines could be identified even if weak), or the spectra of metal- and dust-rich GRB absorbers (in which a significant column density of H$_2$ could to be present).

\section{Conclusion}
\label{sec:conc}

We have presented optical-to-NIR X-shooter spectroscopy of the afterglow of GRB~120815A at $z = 2.36$, supplemented by optical/NIR photometry from GROND and X-ray data from the \textit{Swift} satellite. The bright afterglow emission, observed through efficient instruments at large telescopes, provides a detailed probe of the physical properties of the inter-stellar medium in a high-redshift star-forming galaxy.

The sightline towards GRB 120815A is characterized by a strong DLA with $\log(N(\hi)/\rm{cm}^{-2}) = 21.95\pm0.10$ and substantial amount of molecular hydrogen with a molecular fraction $f$ of 7\% ($\log f(\rm{H}_2)=-1.14\pm0.15$), characteristic of Galactic diffuse clouds. This presents only the second unambiguous detection of H$_2$ in a GRB-DLA, and the first for a GRB-DLA with properties very similar to an average GRB sightline. In addition, we detect vibrationally-excited states of H$_2$, which opens a second route for positive searches of molecular gas in GRB afterglow spectra. 

Our measurements of DLA metallicity ($\rm{[Zn/H]} = -1.15 \pm 0.12$), relative abundance ($\rm{[Zn/Fe]} = 1.01 \pm 0.10$) and visual extinction ($A_V \lesssim 0.15\,\rm{mag}$) are common among GRB-DLAs, and likely typical for the population of GRB-DLAs in general. GRB~120815A thus stands in marked contrast to the metal-rich, H$_2$-bearing DLA of GRB~080607. This illustrates that H$_2$ is present in at least a fraction of the average GRB-DLAs systems. The detection rate of molecular gas in GRB-DLAs could increase once similar afterglows are observed routinely at higher spectral resolution. 

Similar to many other physical properties probed by GRBs or their hosts, such as the distributions of galaxy brightness, mass, star-formation rate \citep{2011arXiv1108.0674K, 2012ApJ...756..187H, 2013arXiv1301.5903P}, galaxy color \citep{2012A&A...545A..77R}, Ly$\alpha$ emission \citep{2012ApJ...756...25M}, dust-reddening \citep{2009ApJ...693.1484C, 2011A&A...526A..30G}, soft X-ray absorption \citep[e.g.,][]{2009ApJS..185..526F, 2010MNRAS.402.2429C, 2012ApJ...758...46K} or metallicity \citep{2009ApJ...691L..27P}, selection effects play an important role when studying H$_2$ in GRB-DLAs. 

A larger sample of afterglows observed in a similar way as presented in this work (Fynbo et al., in preparation) will enable further progress in this field and deeper insights with respect to the statistical presence and properties of molecules along GRB sightlines. Additional detections of molecules in GRB-DLAs would allow detailed individual as well as statistical studies, and, coupled with host follow-up and sub-mm spectroscopy, provide unprecedented insights into the properties of molecular gas and the process and conditions of star-formation at $z\sim2$ and above. 

\begin{acknowledgements}
We acknowledge very helpful comments and a very timely report from the anonymous referee that helped to improve the quality of the manuscript, and thank S. Savaglio for their insightful and constructive comments. TK acknowledges support by the European Commission under the Marie Curie Intra-European Fellowship Programme in FP7. JPUF acknowledges support from the ERC-StG grant EGGS-278202. AdUP acknowledges support by the European Commission under the Marie Curie Career Integration Grant programme (FP7-PEOPLE-2012-CIG 322307). The Dark Cosmology Centre is funded by the Danish National Research Foundation. PS acknowledges support through the Sofja Kovalevskaja Award from the Alexander von Humboldt Foundation of Germany.  S. Schmidl acknowledges support by the Th\"uringer Ministerium f\"ur Bildung, Wissenschaft und Kultur under FKZ 12010-514. TK acknowledges the use of the free software VPFit, written by Bob Carswell and John Webb. Part of the funding for GROND (both hardware as well as personnel) was generously granted from the Leibniz-Prize to Prof. G. Hasinger (DFG grant HA 1850/28-1). This work made use of data supplied by the UK {\it Swift} Science Data Centre at the University of Leicester. Finally, we acknowledge expert support from the ESO staff at the Paranal and La Silla observatories in obtaining these target of opportunity data.

\end{acknowledgements}

\newpage

\begin{appendix}

\section{GROND photometry and multi-color light curves}

{The temporal evolution of the optical/NIR photometric and X-ray data was fitted simultaneously with a two-fold broken power-law, connected smoothly at the break times following \citet{2009A&A...508..593K}. The functional form of this empirical fit is provided for example in \citet{2011A&A...526A..23S}. The light curve shape is characterized by a shallow, early rise with an index $\alpha_1 = -0.18\pm0.02$, where the sign follows the convention that $F_\nu(t) \propto t^{-\alpha}$. The optical/NIR afterglow peaks at $t_1 = 440\pm30$~s after the BAT trigger time. Afterwards the light curve decays with an index of $\alpha_2= 0.52 \pm 0.01$, before another breaking to a steeper decay of $\alpha_3= 0.86 \pm 0.03$  at $t_2 = 4.3_{-0.6}^{+0.9}$~ks. The light curve is reasonably well fitted with this phenomenological model ($\chi^2 = 261$ for 242 degrees of freedom), and no chromatic evolution is apparent within our data set. The overall light curve behavior is reminiscent of GRB~080710 \citep{2009A&A...508..593K}, in particular the achromatic and shallow, early rise of the light curve in combination with the soft $\gamma$-ray emission detected by BAT \citep{2012GCN..13652...1M}.}

\begin{figure*}
\begin{center}
\includegraphics[angle=0, width=0.5\columnwidth]{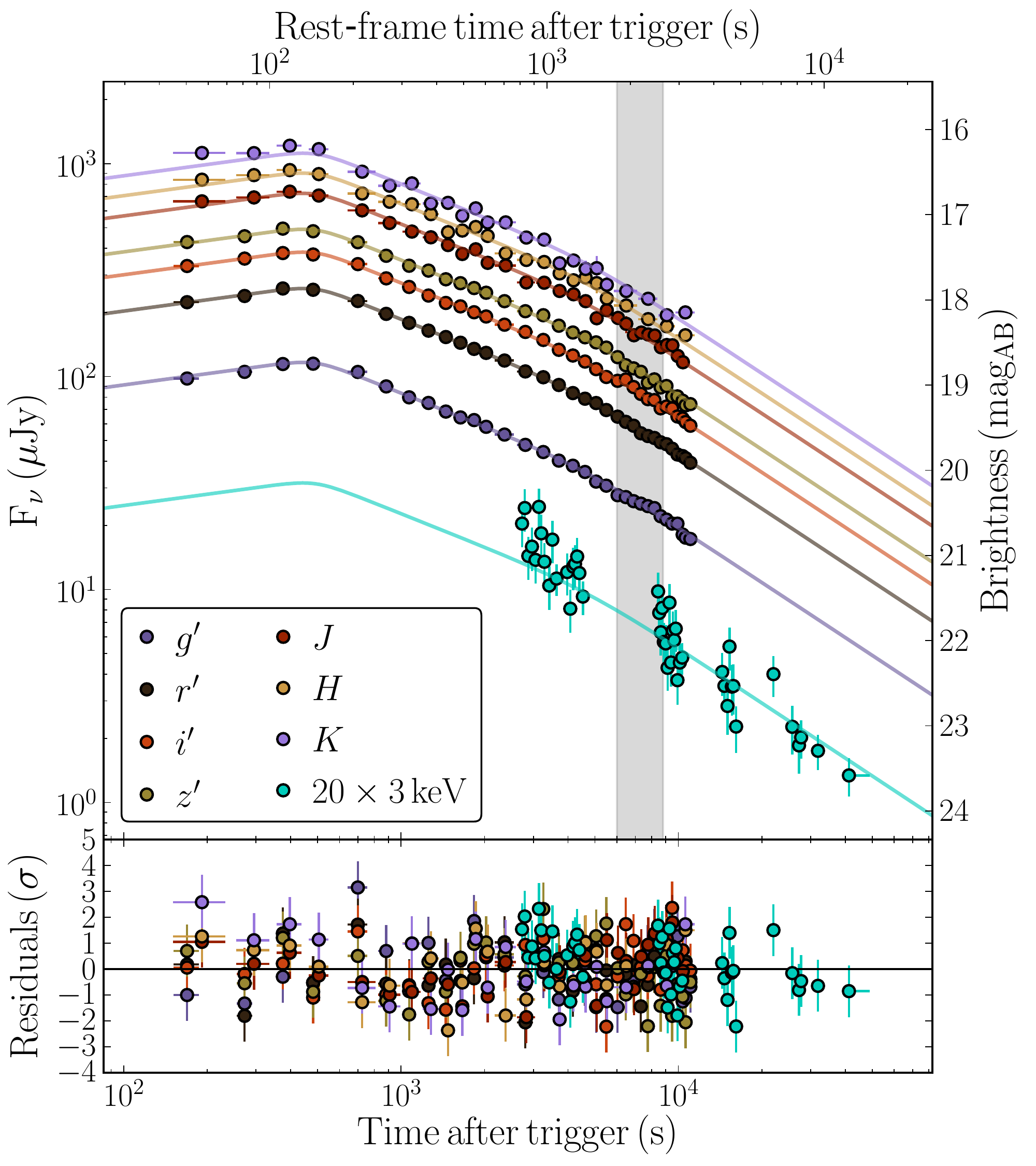}
\caption{GROND optical/NIR and XRT 0.3-10~keV light curves of the afterglow of GRB~120815A in the top panel. The data were fitted with a three-segment, smoothly broken power-law, with residuals shown in the bottom panel. The grey shaded area represents the time interval in which the X-shooter spectroscopy was obtained. X-ray data were converted into a flux density at 3~keV using a spectral index of $\beta= 0.78$, and scaled by a factor of 20 to enhance clarity.}
\label{fig:lc}
\end{center}
\end{figure*}

All magnitudes in Figure~\ref{fig:lc} and Tables~\ref{tab:grizphot}, \ref{tab:JHKphot} are in the AB system and uncorrected for the expected Galactic foreground extinction, corresponding to a reddening of ${E}_{B-V}=0.10\,\rm{mag}$ \citep{2011ApJ...737..103S}.

\begin{table*}
\begin{center}
\caption{Optical magnitudes of the afterglow of GRB~120815A as measured by GROND\label{tab:grizphot}}

\begin{tabular}{ccccccc}
\hline
{Time}    & {Exposure$^{\rm{(a)}}$}  &  \multicolumn{4}{c}{Brightness$^{\rm{(b)}}$} \\ 
{}    &  {}  &  {$g'$-band}   & {$r'$-band}  & {$i'$-band} & {$z'$-band} \\
{(ks after $T_0$)$^{\rm{(c)}}$} & {(s)}  &{(mag$_{\rm{AB}}$)}  & {(mag$_{\rm{AB}}$)} & {(mag$_{\rm{AB}}$)} & {(mag$_{\rm{AB}}$)} \\
\hline
\hline

 0.169	&	35	&	18.92	$\pm$	0.03	&	18.03	$\pm$	0.01	&	17.60	$\pm$	0.02	&	17.32	$\pm$	0.02	\\
0.273	&	35	&	18.85	$\pm$	0.03	&	17.95	$\pm$	0.01	&	17.52	$\pm$	0.02	&	17.25	$\pm$	0.02	\\
0.375	&	35	&	18.75	$\pm$	0.03	&	17.87	$\pm$	0.01	&	17.45	$\pm$	0.02	&	17.16	$\pm$	0.02	\\
0.482	&	35	&	18.75	$\pm$	0.02	&	17.88	$\pm$	0.02	&	17.47	$\pm$	0.02	&	17.19	$\pm$	0.02	\\
0.698	&	115	&	18.85	$\pm$	0.02	&	18.01	$\pm$	0.01	&	17.58	$\pm$	0.01	&	17.32	$\pm$	0.02	\\
0.884	&	115	&	19.02	$\pm$	0.02	&	18.16	$\pm$	0.01	&	17.75	$\pm$	0.02	&	17.48	$\pm$	0.02	\\
1.069	&	115	&	19.14	$\pm$	0.01	&	18.27	$\pm$	0.01	&	17.85	$\pm$	0.01	&	17.60	$\pm$	0.02	\\
1.256	&	115	&	19.21	$\pm$	0.01	&	18.36	$\pm$	0.01	&	17.95	$\pm$	0.02	&	17.66	$\pm$	0.02	\\
1.451	&	115	&	19.31	$\pm$	0.01	&	18.44	$\pm$	0.01	&	18.04	$\pm$	0.02	&	17.76	$\pm$	0.02	\\
1.638	&	115	&	19.38	$\pm$	0.01	&	18.50	$\pm$	0.01	&	18.08	$\pm$	0.01	&	17.80	$\pm$	0.02	\\
1.834	&	115	&	19.41	$\pm$	0.01	&	18.57	$\pm$	0.01	&	18.14	$\pm$	0.02	&	17.86	$\pm$	0.02	\\
2.027	&	115	&	19.49	$\pm$	0.01	&	18.62	$\pm$	0.01	&	18.19	$\pm$	0.02	&	17.92	$\pm$	0.02	\\
2.361	&	375	&	19.58	$\pm$	0.01	&	18.71	$\pm$	0.01	&	18.29	$\pm$	0.01	&	18.02	$\pm$	0.02	\\
2.807	&	375	&	19.70	$\pm$	0.01	&	18.84	$\pm$	0.01	&	18.38	$\pm$	0.02	&	18.13	$\pm$	0.02	\\
3.258	&	375	&	19.79	$\pm$	0.02	&	18.91	$\pm$	0.01	&	18.47	$\pm$	0.02	&	18.18	$\pm$	0.02	\\
3.707	&	375	&	19.89	$\pm$	0.01	&	19.00	$\pm$	0.01	&	18.59	$\pm$	0.03	&	18.30	$\pm$	0.02	\\
4.165	&	375	&	19.95	$\pm$	0.01	&	19.09	$\pm$	0.01	&	18.65	$\pm$	0.02	&	18.38	$\pm$	0.02	\\
4.607	&	375	&	20.02	$\pm$	0.02	&	19.16	$\pm$	0.01	&	18.72	$\pm$	0.02	&	18.44	$\pm$	0.02	\\
5.052	&	375	&	20.13	$\pm$	0.02	&	19.21	$\pm$	0.02	&	18.81	$\pm$	0.02	&	18.50	$\pm$	0.03	\\
5.498	&	375	&	20.18	$\pm$	0.03	&	19.29	$\pm$	0.01	&	18.91	$\pm$	0.02	&	18.56	$\pm$	0.02	\\
6.037	&	$4 \times 35$	&	20.29	$\pm$	0.02	&	19.37	$\pm$	0.03	&	18.95	$\pm$	0.03	&	18.67	$\pm$	0.03	\\
6.465	&	$4 \times 35$	&	20.32	$\pm$	0.04	&	19.43	$\pm$	0.03	&	18.93	$\pm$	0.04	&	18.77	$\pm$	0.04	\\
6.895	&	$4 \times 35$	&	20.36	$\pm$	0.05	&	19.48	$\pm$	0.02	&	19.02	$\pm$	0.03	&	18.80	$\pm$	0.03	\\
7.326	&	$4 \times 35$	&	20.39	$\pm$	0.03	&	19.57	$\pm$	0.02	&	19.10	$\pm$	0.03	&	18.84	$\pm$	0.04	\\
7.763	&	$4 \times 35$	&	20.42	$\pm$	0.04	&	19.60	$\pm$	0.03	&	19.17	$\pm$	0.03	&	18.97	$\pm$	0.04	\\
8.196	&	$4 \times 35$	&	20.44	$\pm$	0.05	&	19.62	$\pm$	0.02	&	19.18	$\pm$	0.03	&	18.93	$\pm$	0.04	\\
8.631	&	$4 \times 35$	&	20.54	$\pm$	0.04	&	19.67	$\pm$	0.02	&	19.27	$\pm$	0.03	&	19.03	$\pm$	0.03	\\
9.062	&	$4 \times 35$	&	20.58	$\pm$	0.04	&	19.69	$\pm$	0.02	&	19.25	$\pm$	0.03	&	19.01	$\pm$	0.03	\\
9.497	&	$4 \times 35$	&	20.63	$\pm$	0.04	&	19.75	$\pm$	0.02	&	19.27	$\pm$	0.03	&	19.13	$\pm$	0.04	\\
9.927	&	$4 \times 35$	&	20.63	$\pm$	0.04	&	19.82	$\pm$	0.02	&	19.37	$\pm$	0.03	&	19.13	$\pm$	0.03	\\
10.363	&	$4 \times 35$	&	20.75	$\pm$	0.03	&	19.83	$\pm$	0.03	&	19.39	$\pm$	0.03	&	19.18	$\pm$	0.04	\\
10.623	&	$4 \times 35$	&	20.79	$\pm$	0.04	&	19.86	$\pm$	0.02	&	19.43	$\pm$	0.03	&	19.24	$\pm$	0.04	\\
11.053	&	$4 \times 35$	&	20.81	$\pm$	0.06	&	19.91	$\pm$	0.03	&	19.48	$\pm$	0.04	&	19.22	$\pm$	0.04	\\

\hline
\hline
\end{tabular} 
\end{center}

\noindent{
$^{\rm{(a)}}$ Integration time of the individual image. Stacked images are given as number of images times exposure of an individual image \newline
$^{\rm{(b)}}$ All magnitudes in this table are in the AB system and uncorrected for Galactic foreground extinction. The quoted error is statistical only, and there is an additional systematic error in the absolute photometric calibration, which is estimated to be around $0.04$~mag in \griz.\newline
$^{\rm{(c)}}$ $T_0$ is set as the time of the \textit{Swift}/BAT trigger, i.e., 2012-08-15 02:13:58 UT.}
\end{table*}

\begin{table*} 
\begin{center}
\caption{Near-infrared magnitudes of the afterglow of GRB~120815A as measured by GROND\label{tab:JHKphot}}
\begin{tabular}{cccccc}
\hline 
{Time}    & {Exposure$^{\rm{(a)}}$}  &  \multicolumn{3}{c}{Brightness$^{\rm{(b)}}$} \\ 
{}    &  {}  &  {$J$-band}   & {$H$-band}  & {$K_{\rm{s}}$-band} \\
{(ks after $T_0$)$^{\rm{(c)}}$} & {(s)}  &{(mag$_{\rm{AB}}$)}  & {(mag$_{\rm{AB}}$)}  & {(mag$_{\rm{AB}}$)} \\
\hline
\hline
0.191	&	$6 \times 10$	&	16.84	$\pm$	0.04	&	16.59	$\pm$	0.05	&	16.28	$\pm$	0.05	\\
0.295	&	$6 \times 10$	&	16.80	$\pm$	0.04	&	16.54	$\pm$	0.05	&	16.28	$\pm$	0.05	\\
0.397	&	$6 \times 10$	&	16.73	$\pm$	0.04	&	16.48	$\pm$	0.05	&	16.19	$\pm$	0.05	\\
0.505	&	$6 \times 10$	&	16.78	$\pm$	0.04	&	16.52	$\pm$	0.05	&	16.23	$\pm$	0.06	\\
0.724	&	$12 \times 10$	&	16.95	$\pm$	0.04	&	16.75	$\pm$	0.05	&	16.50	$\pm$	0.05	\\
0.909	&	$12 \times 10$	&	17.10	$\pm$	0.04	&	16.85	$\pm$	0.05	&	16.66	$\pm$	0.05	\\
1.094	&	$12 \times 10$	&	17.20	$\pm$	0.04	&	16.88	$\pm$	0.05	&	16.63	$\pm$	0.06	\\
1.281	&	$12 \times 10$	&	17.27	$\pm$	0.04	&	16.99	$\pm$	0.05	&	16.87	$\pm$	0.06	\\
1.477	&	$12 \times 10$	&	17.36	$\pm$	0.04	&	17.21	$\pm$	0.05	&	16.86	$\pm$	0.06	\\
1.663	&	$12 \times 10$	&	17.46	$\pm$	0.05	&	17.19	$\pm$	0.05	&	17.01	$\pm$	0.06	\\
1.859	&	$12 \times 10$	&	17.41	$\pm$	0.04	&	17.14	$\pm$	0.05	&	16.93	$\pm$	0.06	\\
2.053	&	$12 \times 10$	&	17.57	$\pm$	0.04	&	17.25	$\pm$	0.05	&	17.09	$\pm$	0.06	\\
2.384	&	$30 \times 10$	&	17.60	$\pm$	0.04	&	17.45	$\pm$	0.05	&	17.09	$\pm$	0.05	\\
2.830	&	$30 \times 10$	&	17.80	$\pm$	0.05	&	17.53	$\pm$	0.05	&	17.27	$\pm$	0.06	\\
3.282	&	$30 \times 10$	&	17.80	$\pm$	0.05	&	17.56	$\pm$	0.05	&	17.29	$\pm$	0.07	\\
3.732	&	$30 \times 10$	&	17.90	$\pm$	0.05	&	17.69	$\pm$	0.06	&	17.57	$\pm$	0.09	\\
4.188	&	$30 \times 10$	&	17.93	$\pm$	0.05	&	17.77	$\pm$	0.08	&	17.54	$\pm$	0.07	\\
4.630	&	$30 \times 10$	&	18.02	$\pm$	0.05	&	17.74	$\pm$	0.07	&	17.64	$\pm$	0.10	\\
5.076	&	$30 \times 10$	&	18.21	$\pm$	0.08	&	17.80	$\pm$	0.11	&	17.63	$\pm$	0.17	\\
5.521	&	$30 \times 10$	&	18.12	$\pm$	0.05	&	17.98	$\pm$	0.07	&	17.82	$\pm$	0.10	\\
6.060	&	$24 \times 10$	&	18.21	$\pm$	0.07	&	\nodata			&	\nodata			\\
6.487	&	$24 \times 10$	&	18.28	$\pm$	0.10	&	\nodata			&	\nodata			\\
6.488	&	$72 \times 10$	&	\nodata			&	18.07	$\pm$	0.07	&	17.90	$\pm$	0.09	\\
6.917	&	$24 \times 10$	&	18.42	$\pm$	0.07	&	\nodata			&	\nodata			\\
7.349	&	$24 \times 10$	&	18.38	$\pm$	0.07	&	\nodata			&	\nodata			\\
7.786	&	$24 \times 10$	&	18.41	$\pm$	0.07	&	\nodata			&	\nodata			\\
7.786	&	$72 \times 10$	&	\nodata			&	18.23	$\pm$	0.05	&	17.95	$\pm$	0.07	\\
8.218	&	$24 \times 10$	&	18.42	$\pm$	0.07	&	\nodata			&	\nodata			\\
8.653	&	$24 \times 10$	&	18.56	$\pm$	0.07	&	\nodata			&	\nodata			\\
9.085	&	$24 \times 10$	&	18.53	$\pm$	0.06	&	\nodata			&	\nodata			\\
9.087	&	$72 \times 10$	&	\nodata			&	18.31	$\pm$	0.05	&	18.15	$\pm$	0.06	\\
9.519	&	$24 \times 10$	&	18.53	$\pm$	0.06	&	\nodata			&	\nodata			\\
9.950	&	$24 \times 10$	&	18.66	$\pm$	0.07	&	\nodata			&	\nodata			\\
10.386	&	$24 \times 10$	&	18.73	$\pm$	0.07	&	\nodata			&	\nodata			\\
10.601	&	$96 \times 10$	&	\nodata			&	18.42	$\pm$	0.05	&	18.11	$\pm$	0.06	\\
10.823	&	$24 \times 10$	&	18.65	$\pm$	0.07	&	\nodata			&	\nodata			\\
11.251	&	$24 \times 10$	&	18.75	$\pm$	0.08	&	\nodata			&	\nodata			\\

\hline
\hline
\end{tabular} 
\end{center}
 
\noindent{
$^{\rm{(a)}}$ Integration time of the individual image. Stacked images are given as number of images times exposure of an individual image \newline
$^{\rm{(b)}}$ All magnitudes in this table are in the AB system and uncorrected for Galactic foreground extinction. The quoted error is statistical only, and there is an additional systematic error in the absolute photometric calibration, which is estimated to be around 0.06~mag in $JH$ and 0.08~mag in $K_{\rm{s}}$.\newline
$^{\rm{(c)}}$ $T_0$ is set as the time of the \textit{Swift}/BAT trigger, i.e., 2012-08-15 02:13:58 UT.}
\end{table*}

\section{X-shooter Spectra of the Afterglow of GRB 120815A}

\begin{figure*}
\begin{center}
\includegraphics[angle=0, width=0.9\columnwidth]{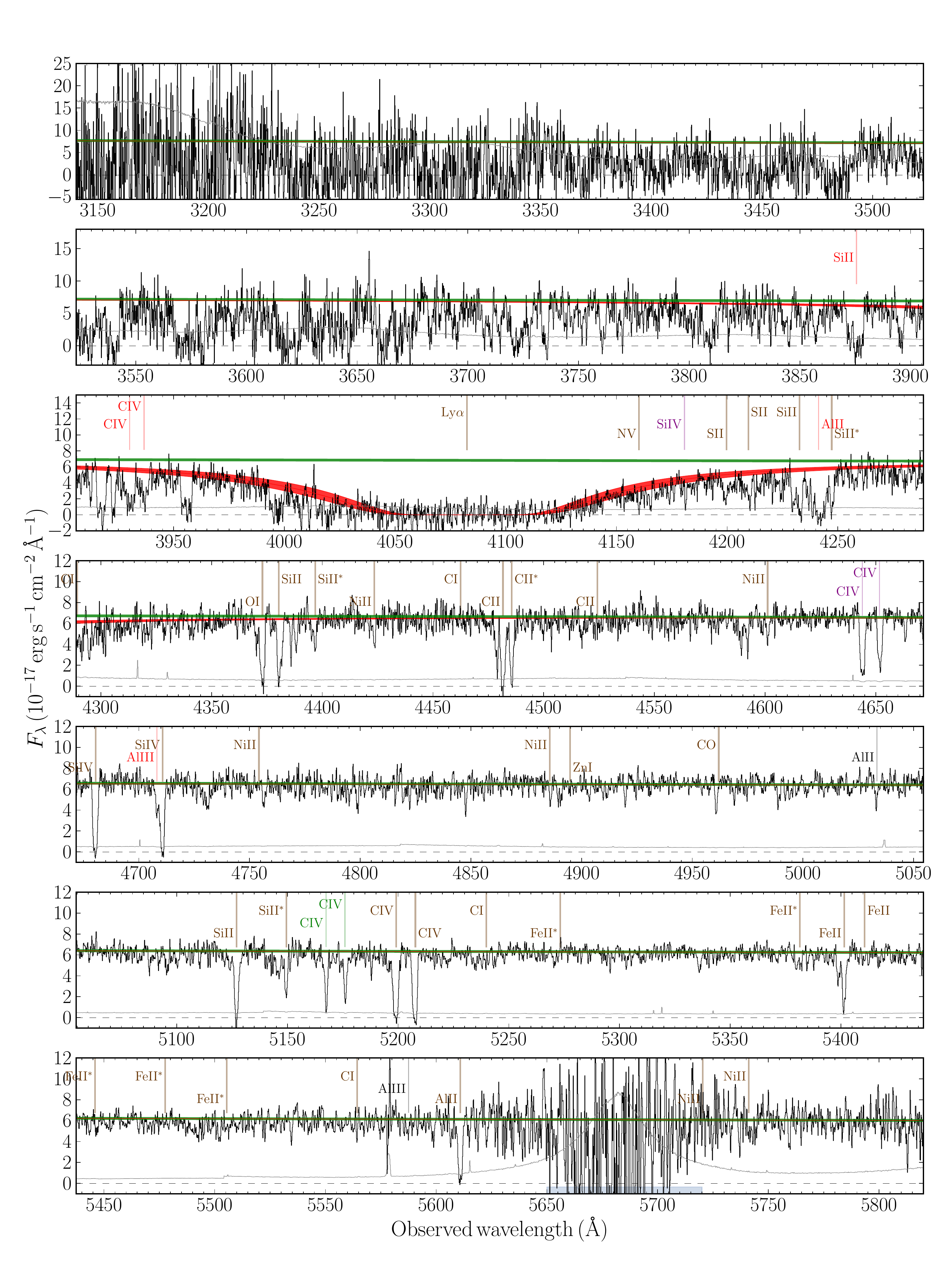}
\caption{X-shooter UVB-arm spectrum of the afterglow of GRB 120815A. Black lines show the spectral data, grey lines the noise level, the green line is the GRB afterglow model, and the red-line shows the DLA modeling. The position of absorption lines that are typically associated with GRB-DLAs \citep[taken from][]{2011ApJ...727...73C} are indicated by brown lines and ions. Individual lines associated with intervening absorbers are labeled with colored lines and labels.}
\label{uvbspec}
\end{center}
\end{figure*}

\begin{figure*}
\begin{center}
\includegraphics[angle=0, width=0.9\columnwidth]{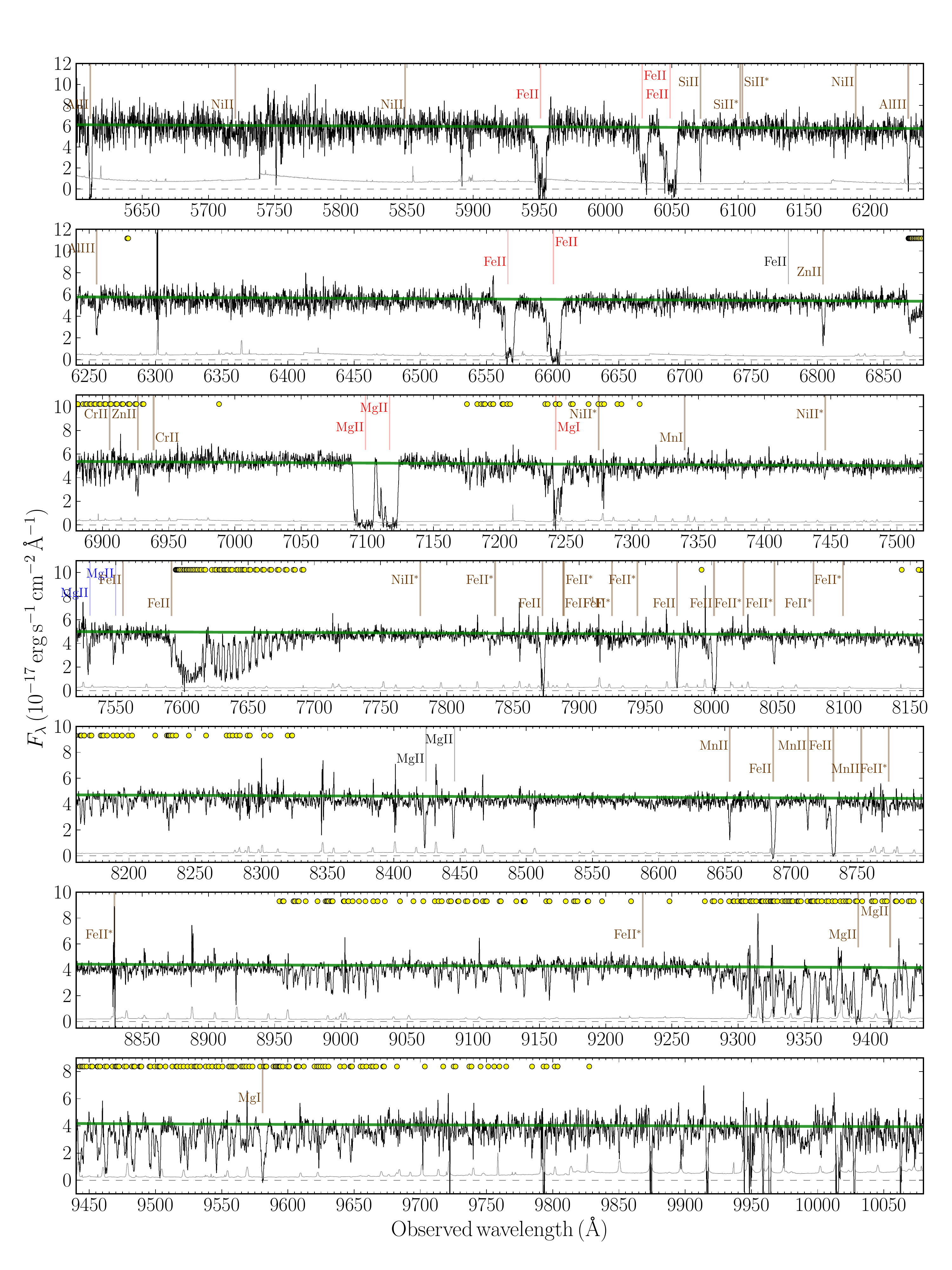}
\caption{X-shooter VIS-arm spectrum of the afterglow of GRB 120815A. Lines and labels are the same as in Figure~\ref{uvbspec}. Additionally, strong telluric lines are indicated by circles in the upper part of each panel.}
\label{visspec}
\end{center}
\end{figure*}

\begin{figure*}
\begin{center}
\includegraphics[angle=0, width=0.9\columnwidth]{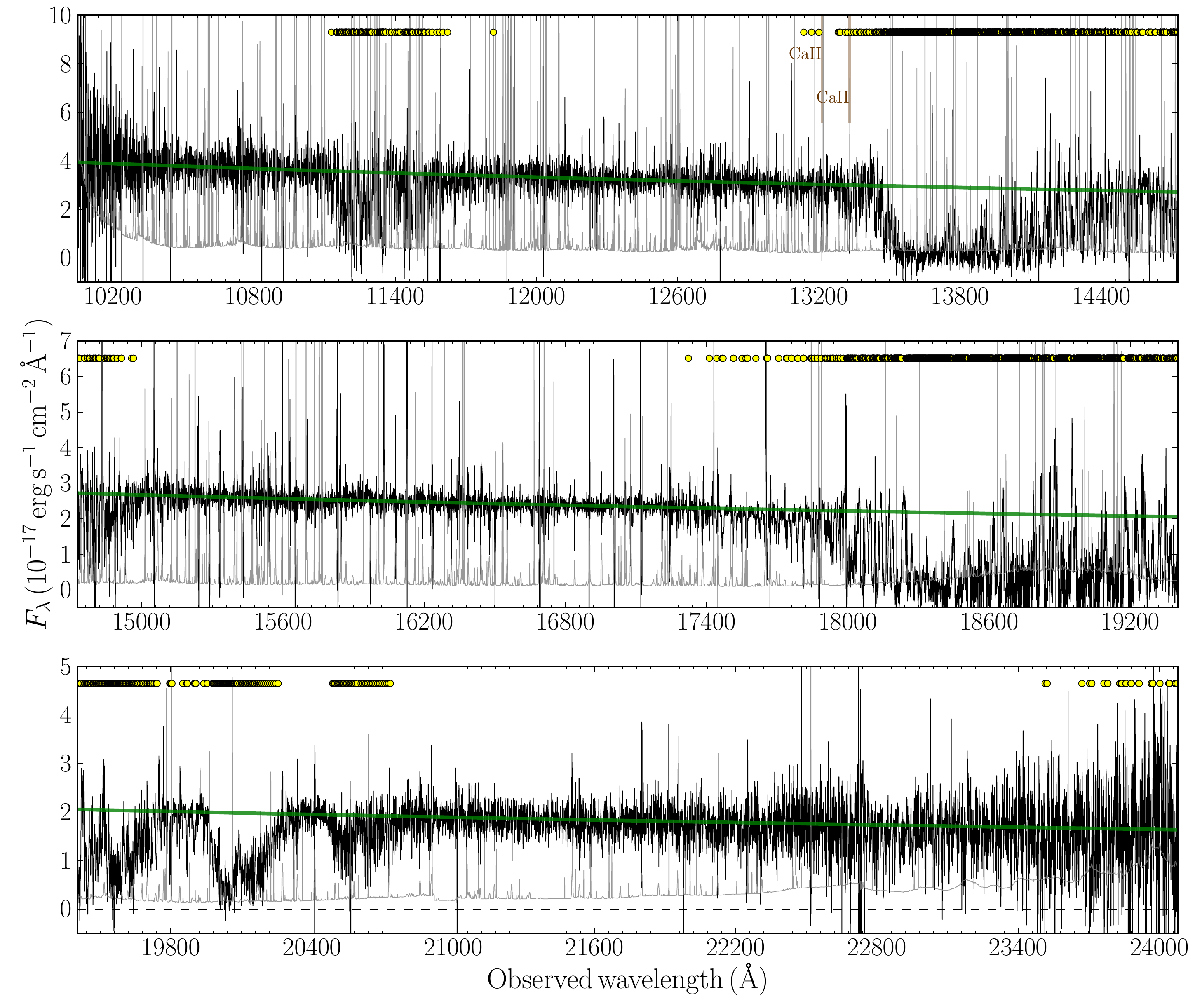}
\caption{X-shooter NIR-arm spectrum of the afterglow of GRB 120815A. Lines and labels are the same as in Figure~\ref{visspec}.}
\label{nirspec}
\end{center}
\end{figure*}

\begin{figure*}
\includegraphics[angle=0, width=0.9\columnwidth]{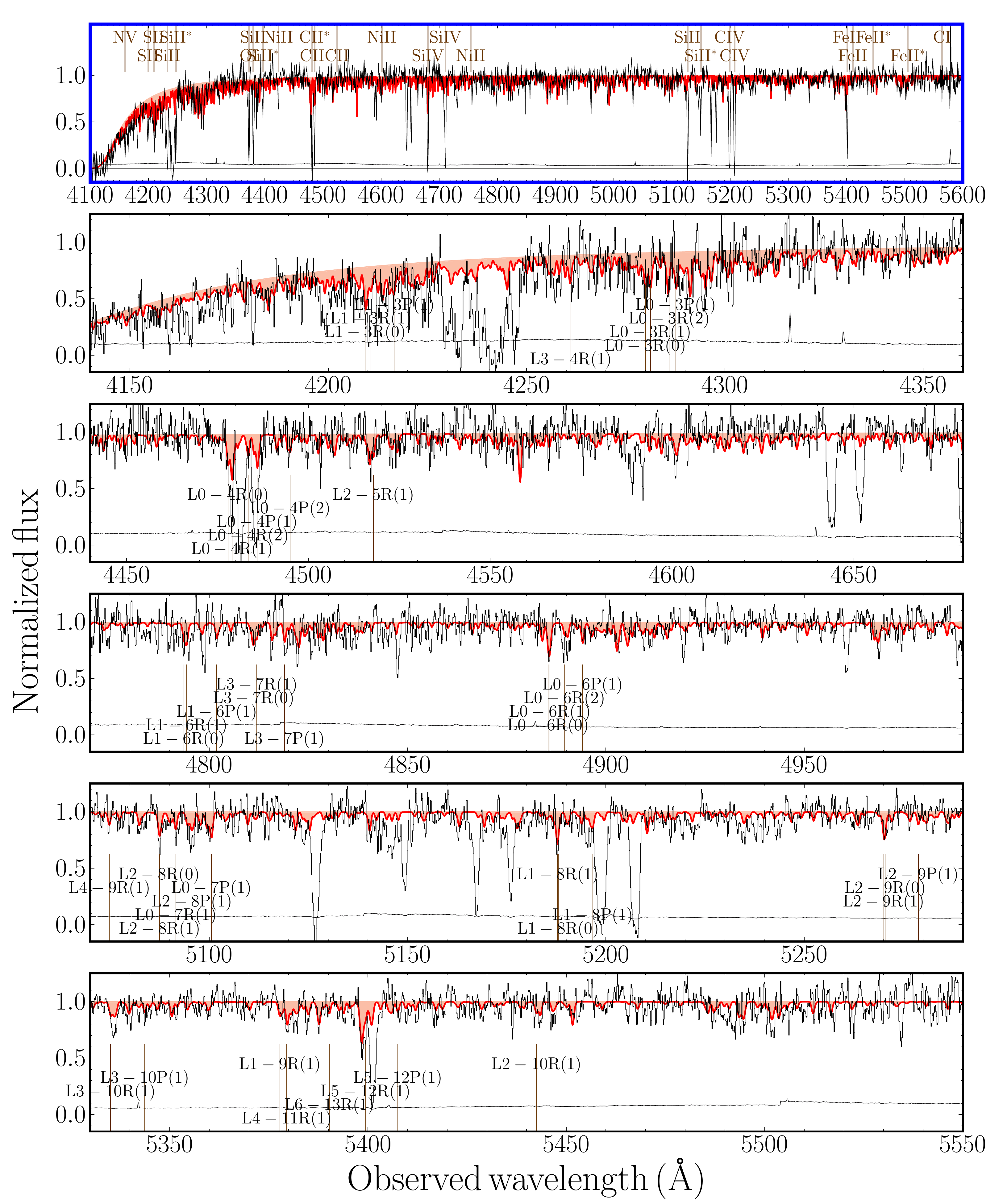}
\caption{X-shooter spectrum between 4100 and 5600~\AA~showing the presence of H$_2^*$ lines. The uppermost, blue-framed panel shows an overview over the whole spectral range, while the lower 5 panels zoom in of 180~\AA~each. We selected regions where most of the absorption bands of H$_2^*$ are located. Light-grey lines always show the normalized spectrum, while dark-grey lines indicate the error spectrum. Red lines denote the best-fit H$_2^*$ model. In the top panel, we also mark prominent metal absorption lines previously detected in GRB-DLAs \citep{2011ApJ...727...73C}. In the lower panels, several individual H$_2^*$ transitions are identified using standard nomenclature with lower and upper vibrational and rotational quantum numbers.}
\label{fig:h2*mod}
\end{figure*}

\end{appendix}
\end{document}